\documentclass[aps, prd, superscriptaddress, nofootinbib, preprintnumbers]{revtex4}
\usepackage{graphics}
\usepackage{epsfig,amsfonts,amssymb,amsmath}
\usepackage{epstopdf}
\usepackage{latexsym}
\usepackage{colordvi}
\usepackage{refstyle}
\usepackage{float}
\newcommand{\be}{\begin{equation}}
\newcommand{\ee}{\end{equation}}
\newcommand{\bea}{\begin{eqnarray}}
\newcommand{\eea}{\end{eqnarray}}
\begin{document}

\title{DC conductivity with external magnetic field in hyperscaling violating geometry}
\author{Neha Bhatnagar}
\email{bhtngr.neha@gmail.com}
\author{Sanjay Siwach}%
 \email{sksiwach@hotmail.com}
\affiliation{%
 Department of Physics, Banaras Hindu University,\\Varanasi-221005, India\\
}%

\begin{abstract}
We investigate the holographic DC conductivity of (2+1) dimensional systems while considering hyperscaling violating geometry in bulk. We consider Einstein-Maxwell-Dilaton system with two gauge fields and Liouville type potential for dilaton. We also consider axionic fields in bulk to introduce momentum relaxation in the system. We apply an external magnetic field to study the response of the system and obtain analytic expressions for DC conductivity, Hall angle and (thermo)electric conductivity.

\end{abstract}


\maketitle

\section{\label{sec:level1}Introduction}
String theory provides us a valuable tool to investigate strongly coupled gauge theories using AdS/CFT duality(holography)\cite{Maldacena:1997re,Gubser:1998bc,Witten:1998qj}. The duality establishes a connection between strongly coupled gauge theory in d-dimensions on the boundary and its weakly coupled gravity dual in (d+1)-dimensional bulk spacetime. Many important phenomenon of nuclear physics, condensed matter physics and high $T_c$ superconductors are being explored using this duality \cite{Son:2002sd, Erlich:2005qh,DaRold:2005zs,Karch:2006pv,Hartnoll:2008kx,Jain:2009bi}.

Recently, considerable interest is seen to study realistic condensed matter systems using holography. The investigation of different phases of strongly coupled systems requires new holographic models. Significant results have been obtained in this area after including momentum dissipation term in holographic models. Realistic examples of strongly coupled systems have finite DC conductivity either due to the presence of impurity or as a result of broken translational invariance. One can introduce momentum relaxation in holographic systems through various ways; by introducing impurity in holographic set-up \cite{Hartnoll:2008hs, Andrade:2013gsa} or by introducing spatial source field which breaks translational symmetry \cite{Donos:2014yya, Amoretti:2014ola}. Also the breaking of diffeomorphism invariance using massive gravity term in the bulk theory, results in finite DC conductivity \cite{Davison:2013jba,Blake:2013bqa,Vegh:2013sk,Amoretti:2014mma,Baggioli:2014roa}.

Further, efforts are being made to study strongly coupled condensed matter systems near quantum critical points using holography. These critical points are realized in the boundary system by opting for anisotropic scaling between temporal and spatial directions in the gravity set-up. Although the introduction of anisotropy results in breaking of Lorentz invariance, the metric remains scale invariant. The system with this anisotropy while possessing scaling symmetry is known as Lifshitz-like geometry,
\begin{equation}
\bar{x} \rightarrow \lambda \bar{ x} \qquad  t =\lambda^z t.
\end{equation}  
where $\bar{x}$ is spatial coordinate, t is temporal coordinate and $z >1$ is known as the dynamical critical exponent.
 
Several efforts are being made to realize this type of geometry (Lifshitz-like) in gravity set-up. The most common way is working with Einstein-Maxwell-Dilaton (EMD) theories in the gravity system. This geometry was first introduced by \cite {Kachru:2008yh} and found wide range of application in analyzing thermodynamical and hydrodynamical aspects of strongly coupled systems \cite{Liu:2009dm,Bertoldi:2010ca,Berglund:2011cp,Iizuka:2011hg,Gouteraux:2011ce,Alishahiha:2012cm,Dey:2012tg,Alishahiha:2012qu,Gouteraux:2014hca}.
 Generalized system with warped metric is also used for the detailed studies of realistic condensed matter system \cite{Cremonini:2016avj,Charmousis:2010zz,Ammon:2012je,Kuang:2014pna,Karch:2014mba,Kuang:2015mlf,Wu:2015ajt,Li:2016rcv}.
\begin{equation}
ds^2_{d+2}=r^{\frac{2\theta}{d}}\left(-r^{2z}dt^2+\frac{dr^2}{r^2}+r^2\sum_{i=1}^d dx_i^2\right),
\end{equation}
where $\theta$  is known as hyperscaling violating parameter in d-dimensions.  

In this work, we investigate the DC conductivity and (thermo)electric conductivity of (2+1) dimensional systems with the hyperscaling violating term for Lifshitz-like geometry. Hence, we consider two different gauge fields in gravity set-up, one field will introduce Lifshitz-like geometry while other introduces finite charge density. Linear axionic fields have been introduced in the system to break translational invariance and obtain finite DC conductivity. Introducing an external magnetic field \cite{Donos:2015bxe, Kim:2015wba, Blake:2015ina, Amoretti:2015gna} has given us an edge for detailed study of the holographic system and is the main motivation of our present work. We discuss the dependence of transport on the dynamical exponents in the presence of external magnetic field and multiple gauge fields. To read transport coefficients for boundary theory using correlation function, one can use various approaches \cite{Papadimitriou:2004rz,Iqbal:2008by,Kuperstein:2011fn,Heemskerk:2010hk,Faulkner:2010jy,Kuperstein:2013hqa,Amoretti:2014zha,Ge:2016lyn,Tian:2017vfk,Hartnoll:2007ai,Hartnoll:2007ih, Blake:2014yla,Ge:2016sel,Amoretti:2017xto,Cremonini:2017qwq,Amoretti:2016cad}.
In this work, we have simplified the calculation while using Wilsonian RG(renormalization group) flow approach\cite{Matsuo:2011fk}. The advantage of this approach is that second order coupled differential equations reduces to first order ordinary differential equations \cite{Kuperstein:2011fn, Kuperstein:2013hqa, Tian:2017vfk}. We have studied these RG flow equations and extract the transport coefficients of the boundary theory.

The paper is organized as follows. We introduce basic holographic set-up of Lifshitz-like geometry with hyperscaling violation term in the next section. We use Wilsonian RG flow approach to study DC conductivity and (thermo)electric conductivity for the system in the presence of the external magnetic field in the following section. We discuss the dependence of transport coefficients on magnetic field and strength of momentum relaxation  by various plots. The temperature dependence of various counductivities is also investigated. The concluding section contains the detailed discussion and summary of the work done. 

\section{EMD system}
Let us consider the EMD system with two gauge fields and axionic fields. The gravity action is given by,
\begin{equation}
S=\int d^4 x \sqrt{-g}\left(R-\frac{1}{2}(\partial \phi)^2+V(\phi)-\frac{1}{4}\sum_{i=1}^{2}e^{\lambda_i \phi}F^2_i -e^{\lambda_3 \phi}\sum_{i=1}^{2}\partial \chi_i^2\right), \label{eq:action}
\end{equation}
where $F_1$ and $ F_2$ are two $U(1)$ gauge fields, the role of first gauge field is to break Lorentz-invariance and introduce Lifshitz-like geometry while second gauge field introduced the charge in the gravitational set-up. $V(\phi)$ is dilaton potential and $\chi_i$ are the axionic fields. In this work, we take the potential of form, $V(\phi)=-2 \Lambda e^{\lambda_0 \phi}$ for further calculation.

The Einstein equation from the above action is given by,
\begin{equation}
R_{\mu \nu}=\frac{1}{2}\partial _{\mu}\phi \partial_{\nu}\phi +\Lambda e^{\lambda_0 \phi}g_{\mu \nu}+\frac{1}{2}e^{\lambda_3 \phi}(\partial_\mu \chi_i \partial_\nu \chi_i)+\sum_{i=1}^2 \frac{1}{2}e^{\lambda_i \phi}(F_{i\mu \rho}F^\rho_{i ~\nu}-\frac{1}{4}F_i^2g_{\mu\nu}).
\end{equation}
The matter fields equations of motion are obtained as,
\begin{eqnarray}
\square\phi= \frac{1}{2}\lambda_3\sum_{i=1}^2(\partial \chi_i)+\frac{1}{4} \sum_{i=1}^2 \lambda_i e^{\lambda_i \phi}F_i^{2} + 2\lambda_0 \Lambda e^{\lambda_0 \phi}, \\
0=\nabla_\mu\left(e^{\lambda_1 \phi} F^{\mu \nu}_1 \right),\quad
0=\nabla_\mu\left(e^{\lambda_2 \phi} F^{\mu \nu}_2 \right), \label{eq:pa1}\\
0=\nabla_\mu\left(e^{\lambda_3 \phi} \nabla^\mu \chi_i\right).
\end{eqnarray}
The metric ansatz for the above action (\ref{eq:action}) is given by Lifshitz-like, hyperscaling violating black-brane solution as in \cite{Gouteraux:2014hca,Alishahiha:2012qu,Li:2016rcv,Cremonini:2016avj},
\begin{equation}
ds^2=r^{\theta}\left(-r^{2z}f(r) dt^2 +\frac{dr^2}{r^2 f(r)}+r^2(dx^2+dy^2)\right).
\end{equation}
The external magnetic field is introduced in the set-up in the given form,
\begin{equation}
A_{2}=A_2(r)dt+Bxdy.
\end{equation}
The axion fields are linear in spatial direction given by, $\chi_1=\alpha x$ and $\chi_2=\alpha y$ where $\alpha$ is considered as the strength of the momentum dissipation.\\
Using the gravity solution, the parameters of given model are related by, \cite{Cremonini:2016avj}
\begin{eqnarray}
\gamma=\sqrt{(\theta+2)(\theta+2 z-2)}, \qquad \lambda_0=\frac{-\theta}{\gamma},\qquad  \lambda_1=\frac{-(4+\theta)}{\gamma}, \\
\lambda_2=\frac{(\theta+2 z-2)}{\gamma}, \qquad \lambda_3=\frac{-\gamma}{\theta+2}, \qquad q_1=\sqrt{2(z-1)(\theta+z+2)}, 
\end{eqnarray}
with $\Lambda=\frac{-1}{2}(\theta+z+1)(\theta+z+2)$ and $\phi =\gamma \log r$. 

From the temporal component of the gauge field equation (\ref{eq:pa1}) we obtain,
\begin{equation}
J_i^t=q_i= r^{-z+3}e^{\lambda_i \phi}(A_i)'_t,
\end{equation}
considering $q_i$ as the charges of two gauge fields. Here the role of $q_1$  is the used to  introduce Lifshitz-scaling whereas $q_2$ is interpreted as the black hole charge.\\
The black hole factor with an external magnetic field (B), mass (m) and charge $(q_2)$ along with axionic strength($\alpha$) is given by,
\begin{equation}
f(r)=1-\frac{m}{r^{\theta+z+2}}+\frac{q_2^2+B^2}{2(\theta+2)(\theta+z)r^{2(\theta +z+1)}}+\frac{\alpha^2}{(\theta+2)(z-2)r^{\theta+2z}}. \label{eq:fs}
\end{equation}
The Hawking temperature of  black hole is obtained from the expression given below,
\begin{equation}
T=\frac{r_h^{z+1}f'(r_h)}{4 \pi}, \label{eq:HT}.
\end{equation}
Thus,
\begin{equation}
T=\frac{z+2+\theta}{4 \pi}r_h^z -\frac{q_2^2}{8 \pi (2+\theta)}\frac{1}{r_h^{z+2+2 \theta}}-\frac{\alpha^2}{4 \pi (2+ \theta)}\frac{1}{r_h^{z+\theta}}. \label{eq:T}
\end{equation}
The constraint from the gravity solution is that every point in space-time follows the null energy condition (NEC) i.e.,
$
T_{\mu \nu}V^{\mu}V^{\nu} \geq 0,
$
where $V^{\mu}$ is the light like vector. Thus, the allowed values of `z' and  `$\theta$' consistent with the gravity solution are \cite{Cremonini:2016avj},
\begin{eqnarray}
(2+\theta)(2z-2+\theta) \geq 0, \\
(z-1)(2+z+\theta) \geq 0.
\end{eqnarray}

Later, we shall see that the consistency of coupled equations demand that $\theta=z-1$ and both the conditions reqire $z\geq 1$. Further at $z=2$, the solution of black hole factor ($\ref{eq:fs}$) is not valid as the last term changes sign. So we shall consider the range $1\leq z <2$, which corresponds to $0\leq\theta<1$. 

To study the response of system, we introduce the following perturbations,
\begin{eqnarray}
\delta A_{1i}(t, r)&=& \int_{-\infty}^{\infty} \frac{ d\omega}{2 \pi}a_{1i}(r)e^{-i \omega t}, \\
\delta A_{2i}(t, r)&=& \int_{-\infty}^{\infty} \frac{ d\omega}{2 \pi}a_{2i}(r)e^{-i \omega t},\\
\delta \chi_i(t, r)&=&\int_{-\infty}^{\infty} \frac{ d\omega}{2 \pi}b_i(r)e^{-i\omega t},\\
\delta g_{ti}(t, r)&=&\int_{-\infty}^{\infty} \frac{ d\omega}{2 \pi}r^{\theta+2} h_{ti}(r)e^{-i \omega t}.
\end{eqnarray}

The coupled linearized equations of motion for the fields are obtained as,
\begin{eqnarray}
\left.\begin{aligned}
0&=(r^{z-3-\theta}f a'_{1i})'+\frac{\omega^2 a_{1i}}{r^{5+z+\theta}f}+q_1h'_{ti},  \label{eq:gauge1}\\
0&=(r^{3z-1+\theta}fa'_{2i})'+ \frac{\omega^2 a_{2i}}{r^{3-z-\theta}f}+q_2h'_{ti}+\epsilon_{ij}\frac{i\omega B h_{tj}}{f r^{3-z-\theta}}\label{eq:eq2} \label{eq:gauge2}, \\
0&=(r^{5-z}fb'_i)'+\frac{\omega^2b_i}{f r^{3z-3}}-\frac{i\omega \alpha h_{ti}}{f r^{3z-3}}. \\
\end{aligned}\right.
\end{eqnarray}
where $\epsilon_{ij}$ is the Levi-Civita tensor. The constraint equation of the set-up is given by,
\begin{equation}
0=\omega r^{5-z+\theta}h'_{ti}+\omega q_1 a_{1i}+\omega q_2 a_{2i}+i \alpha r^{5-z}f b_i'-q_2 B h_{ti}-Bfa'_{z_i}r^{3z-1+\theta}. \label{eq:cons}
\end{equation}
whereas metric perturbation equation is obtained in the given form,
\begin{eqnarray}
0=(r^{5-z+\theta} h'_{ti})'-q_1  r^{z-1-\theta}a'_{1i}-q_2  r^{z-1-\theta}a'_{2i} +\frac{(\alpha^2r^{-\theta-2z+2}+B^2 r^{2z-4})h_{ti} }{f} \nonumber \\+\frac{i \alpha \omega r^{-\theta-2z+2}b_i}{f} +\epsilon_{ij}\frac{i\omega B r^{2z-4}a_{2i}}{f}. \label{eq:metric}
\end{eqnarray}

\section{Holographic Approach}
To study the transport properties of (2+1) dimensional  boundary system, one has to solve coupled equations of motion (\ref{eq:gauge1}), (\ref{eq:cons}) and (\ref{eq:metric}) using the standard procedure as mentioned in \cite{Blake:2014yla,Kim:2015wba}. However in this work we follow the approach introduced by \cite{Matsuo:2011fk} and developed through various studies\cite{Kuperstein:2011fn,Kuperstein:2013hqa,Tian:2017vfk}. Here, DC conductivities are obtained from first order RG flow equations in the near horizon limit. To study the conductivity of a system we simply apply Ohm's Law as given below.
\begin{equation}
J=\sigma E.
\end{equation}
Applying holographic techniques to obtain transport coefficients we use the Onsager relation ($J=\tau X$) in the matrix form as shown,
\begin{equation}
\begin{pmatrix} J_{1i} & J_{1j} \\ J_{2i} & J_{2j}\end{pmatrix} =\tau \begin{pmatrix}X_{1i}& X_{1j}\\ X_{2i} & X_{2j} \end{pmatrix},
\end{equation}
where `$X_i$' are the linear independent  sources  and  `$J_i$' are the responses of system. $\tau$ matrix are the coefficients  evaluated in the near horizon limit. The detailed discussion and application of the formalism is presented in \cite{Matsuo:2011fk,Ge:2016sel,Tian:2017vfk}. Also, we could use the following notation 
\begin{equation}
\begin{Vmatrix}
J_1\\J_2
\end{Vmatrix}=\tau \begin{Vmatrix}X_1 \\X_2 \end{Vmatrix}
\end{equation}
And $\tau =J X^{-1}$ can be expressed as,
\begin{equation}
\tau= \begin{Vmatrix} J_1\\ J_1 \end{Vmatrix} \begin{Vmatrix}X_1 \\X_2 \end{Vmatrix}^{-1}
\end{equation}

The linearized equations of motion (\ref{eq:gauge2}) and (\ref{eq:metric}) can be put in the matrix form as,
\begin{equation}
\tau =
\begin{Vmatrix}
-r^{z-3-\theta} f a_{1i'}\\-r^{3z-1+\theta} fa_{2i}' \\ -r^{5-z}f b_i'\\-r^{5-z+\theta}h_{ti}' \end{Vmatrix} \begin{Vmatrix}
i \omega a_{1i}\\ i \omega a_{2i}\\ i \omega b_{i}\\ i \omega h_{ti} 
\end{Vmatrix}^{-1} \label{eq:tau}
\end{equation}
Now, taking the radial derivative and substituting the equations of motion we get,
\begin{eqnarray}
\tau'=
\begin{Vmatrix}
\frac{\omega^2 a_{1i}}{r^{5+z+\theta}f}+q_1 h'_{ti}\\
 \frac{\omega^2 a_{2i}}{r^{3-z-\theta}f}+q_2 h'_{ti}+\epsilon_{ij}\frac{\omega B h_{tj}}{f r^{3-z-\theta}}\\
 \frac{\omega^2b_i}{f r^{3z-3}}-\frac{i\omega \alpha h_{ti}}{f r^{3z-3}}\\
q_1  r^{z-1-\theta}a'_{1i}+q_2  r^{z-1-\theta}a'_{2i} +\frac{\omega B r^{2z-4} \epsilon_{ij}a_{2j}}{f}-\frac{i \alpha \omega r^{-\theta-2z+2}b_i}{f}-\frac{(\alpha^2r^{-\theta-2z+2}+B^2 r^{2z-4})h_{ti}}{f} 
\end{Vmatrix}
\begin{Vmatrix}
i \omega a_{1i}\\ i \omega a_{2i}\\ i \omega b_{i}\\ i \omega h_{ti} 
\end{Vmatrix}^{-1}\nonumber \\ -\tau 
\begin{pmatrix}
\frac{-i \omega}{r^{z-3-\theta}f}&0&0&0\\
0&\frac{-i \omega}{r^{3z-1+\theta}f}&0&0\\
0&0&\frac{-i \omega}{r^{5-z}f}&0\\
0&0&0&\frac{-i \omega}{r^{5-z+\theta}}
\end{pmatrix} 
\tau \nonumber
\end{eqnarray}
Further simplifying we get,
\begin{eqnarray}
\tau'=
\begin{pmatrix}
\frac{-i \omega}{r^{z+5+\theta}f} &0&0&0\\
0&\frac{-i \omega}{r^{3-z-\theta}f} &0&\frac{i B}{ r^{3-z-\theta}f}\\
0&0&\frac{-i\omega}{ r^{3z-3}f}&-\frac{\alpha}{ r^{3z-3}f} \nonumber \\
0&\frac{-i  B r^{2z-4}}{f}&\frac{-\alpha r^{-\theta-2z+2}}{f}&-\frac{(B^2 r^{2z-4}+\alpha^2r^{-\theta-2z+2})}{i \omega f}\\
\end{pmatrix}\nonumber \\+ 
\begin{pmatrix}
\frac{-q_1}{r^{5-z+\theta}} &0&0&0\\
0&0&0&\frac{-q_2}{r^{5-z+\theta}}\\
0& 0& 0& 0 \\
\frac{-q_1}{r^{-2}f}&\frac{-q_2}{r^{2z+2\theta}f}& 0 & 0\\
\end{pmatrix} 
\tau  -\tau 
\begin{pmatrix}
\frac{-i \omega}{r^{z-3-\theta}f}&0&0&0\\
0&\frac{-i \omega}{r^{3z-1+\theta}f}&0&0\\
0&0&\frac{-i \omega}{r^{5-z}f}&0\\
0&0&0&\frac{-i \omega}{r^{5-z+\theta}}\\
\end{pmatrix}
\tau
\end{eqnarray} 
Multiplying the above metric by  black hole factor `f(r)'  and taking its near horizon limit, where $f(r_h)=0$ we obtain $\tau_h$.\\
Considering the constraint equation (\ref{eq:cons}) in the matrix form as,
\begin{equation}
\begin{pmatrix}   0& 0&0 &0\\ 0& i B & \alpha & -i \omega \\ 0& 0&0 &0\\  0& 0&0 &0 \end{pmatrix} \tau= \begin{pmatrix}  0& 0&0 & 0 \\-q_1 & -q_2 &  0  & \frac{-q_2 B}{\omega} \\ 0& 0&0 &0\\  0& 0&0 &0 \end{pmatrix}
\end{equation}
we obtain,
\begin{eqnarray}
iB \tau_{11}+\alpha \tau_{21}-i \omega\tau_{31}&=&-q_1, \\
iB \tau_{12}+\alpha \tau_{22}-i \omega\tau_{32}&=&-q_2, \\
iB \tau_{13}+\alpha \tau_{23}-i \omega\tau_{33}&=&0, \\
iB \tau_{14}+\alpha \tau_{24}-i \omega\tau_{34}&=&-\frac{q_2 B}{\omega}.
\end{eqnarray}

To maintain the consistency of the equations, we fixed $\theta =z-1$ and the $\tau_h$ matrix is given as,
\begin{equation}
\tau_h =
\begin{pmatrix}
 r_h^{-3-z}&0&0&0\\
0&r_h^{-3+3z}& 0& \frac{-B r_h^{-3+3z}}{\omega}\\
0& 0& r_h^{-2z+4} & \frac{\alpha r_h^{-2z+4}}{i \omega}\\
 \frac{q_1}{i \omega}&\frac{i Br_h^{-3+2z}+q_2}{i \omega}&\frac{r_h^{4-2z}}{i \omega}&-\frac{B^2 r_h^{3z-3}+\alpha^2 r_h^{4-2z}-iq_2B}{\omega^2}\\
\end{pmatrix}
\end{equation}
By substituting $\tau_h$ in equation (\ref{eq:tau}), we find the flow equation in the near horizon limit,
\begin{eqnarray}
\left.\begin{aligned}
(-r_h^{-2}fa_{1i})'&=r_h^{-3-z}i \omega a_{1i}\label{eq:f1},\\
(-r_h^{4z-2}fa_{2i})'&=r_h^{-3+3z}i \omega a_{2i}+\epsilon_{ij}iBr_h^{-3+3z}h_{tj},\\
(-r_h^{5-z}f b_i)'&=r_h^{-2z+4}i\omega b_i+\alpha r_h^{-2z+4}h_{ti} \label{eq:f1},\\
(-r_h^{4}h_{ti})'&=(iBr_h^{-3+3z}+q_2)a_{2i}+r_h^{4-2z}\alpha b_i+q_1a_{1i}-\frac{(iB^2 r_h^{3z-3}+\alpha^2r_h^{4-2z}+q_2B)h_{tj}}{\omega}.
\end{aligned}\right.
\end{eqnarray}
Near horizon limit of equation (\ref{eq:metric}) is given by,
\begin{equation}
(B^2 r_h^{2z-4}+\alpha^2 r_h^{-3z+3})h_{ti}=q_1a_{1i}' +q_2 a_{2i}'+\epsilon_{ij}\omega B a_{2j}r_h^{2z-4}-i \alpha \omega r_h^{-3z+3}b_i.
\end{equation}
Simplifying the above expression while using the flow equations (\ref{eq:f1}) we obtain the expression for the metric perturbation as,
\begin{equation}
h_{ti}|_{r=r_h}=-i \omega\frac{(q_1 r_h^{-z-1})a_{1i} + \epsilon_{ij}(q_2r_h^{-z-1}+i B r_h^{2z-4})a_{2j}}{(B^2 r_h^{2z-4}+\alpha^2 r_h^{-3z+3})-i q_2 Br_h^{-z-1}} -\frac{i \alpha \omega r_h^{-3z+3}b_i}{(B^2 r_h^{2z-4}+\alpha^2r_h^{-3z+3})-i q_2 Br_h^{-z-1}}. \label{eq:mi}
\end{equation}
 
\subsection{DC conductivity}
Using gauge field equation (\ref{eq:gauge1}), we get the expression for conserved currents for first gauge field as,
\begin{equation}\label{eq:cur1}
J_{1i}=-r^{z-3-\theta} f a_{1i}'-q_1h_{ti}.
\end{equation}
On substituting the equation (\ref{eq:f1}) the expression takes the form,
\begin{equation}
J_{1i}=r_h^{-3-z}i \omega a_{1i}-q_1 h_{ti}.
\end{equation}
Now from equation (\ref{eq:mi}) neglecting the second term (i.e. axion perturbation part) we obtain the DC conductivity using,
\begin{equation}
\sigma_{ij} =\frac{\partial J_i}{\partial E_j},\qquad {\text{where}} \quad E_j=i \omega a_j.
\end{equation}
Thus we obtain,
\begin{equation}
\sigma_{xx}^{11}=\sigma_{yy}^{11}= r_h^{-3-z}+\frac{q_1^2 \left(B^2 r_h^{z-5}+\alpha^2 r_h^{-4z+2}\right)}{(B^2 r_h^{2z-4}+\alpha^2 r_h^{-3z+3})^2+B^2q_2^2 r_h^{-2z-2}}, \label{eq:si1}
\end{equation}
Also, we have some mixed terms for DC conductivity, where charges of both the fields effect the conductivity as shown,
\begin{eqnarray}
\sigma_{xx}^{12}&=&\sigma_{yy}^{12}=\frac{q_1 q_2 \alpha^2 r_h^{-4z+2}}{(B^2 r_h^{2z-4}+\alpha^2 r_h^{-3z+3})^2+B^2q_2^2 r_h^{-2z-2}},\\
\sigma_{xy}^{11}&=&-\sigma_{yx}^{11}=\frac{q_1^2 q_2 B r_h^{-2}}{(B^2 r_h^{2z-4}+\alpha^2 r_h^{-3z+3})^2+B^2q_2^2 r_h^{-2z-2}},\\
\sigma_{xy}^{12}&=&-\sigma_{yx}^{12}=q_1 B\frac{B^2 r_h^{4z-8}+\alpha^2 r_h^{-z-1}+q_2^2 r_h^{-2-2z}}{(B^2 r_h^{2z-4}+\alpha^2 r_h^{-3z+3})^2+B^2q_2^2 r_h^{-2z-2}}.
\end{eqnarray}
Since these expression are quite complex to analyse we numerically study the dependence of conductivities on magnetic field and momentum dissipation term for two different values of the dynamical exponent, $z=1$  and $z=4/3$ in Fig.1 to Fig.4. 
\begin{figure}[h!]
\centering
  \includegraphics[width=.45\textwidth]{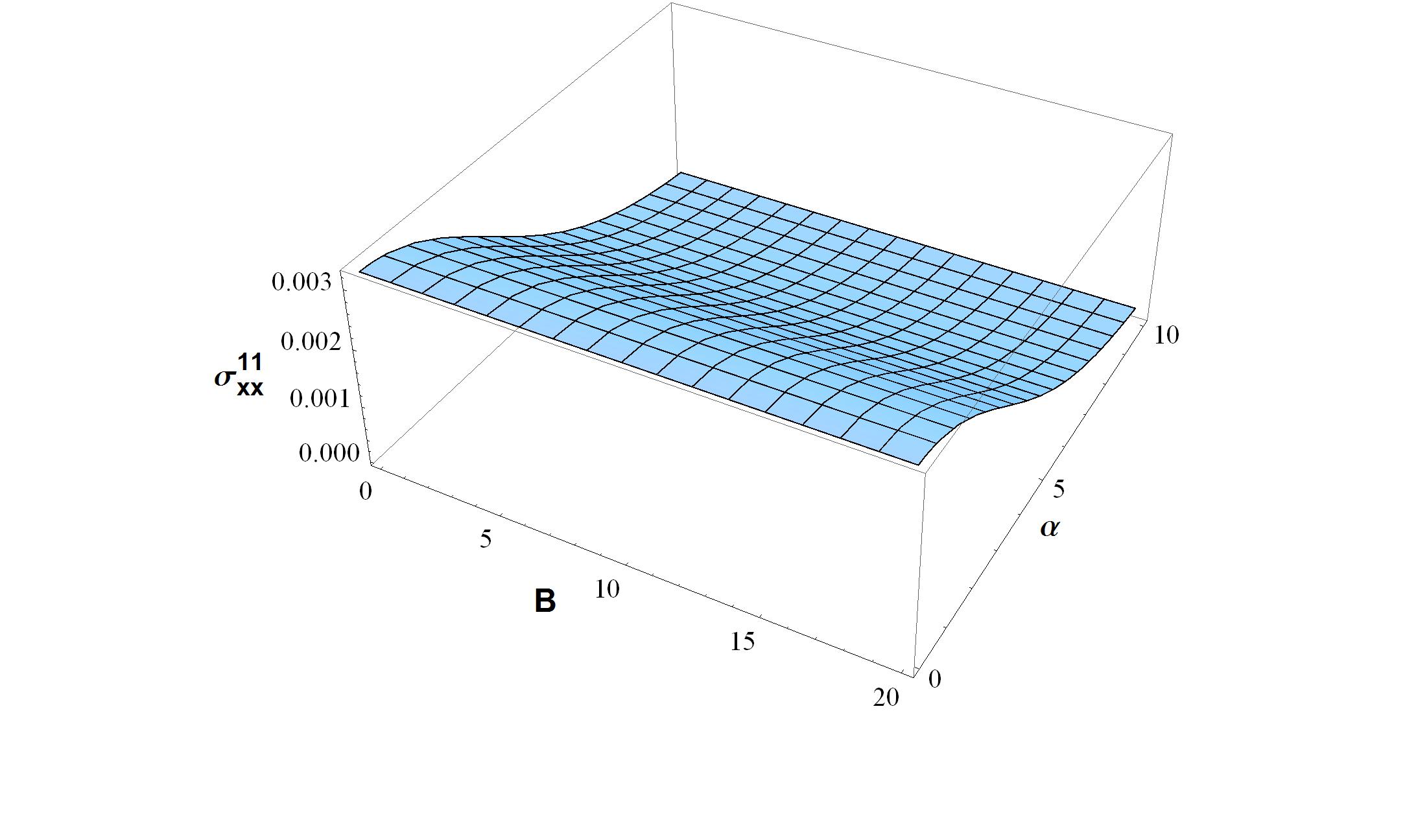}
  \includegraphics[width=.45\textwidth]{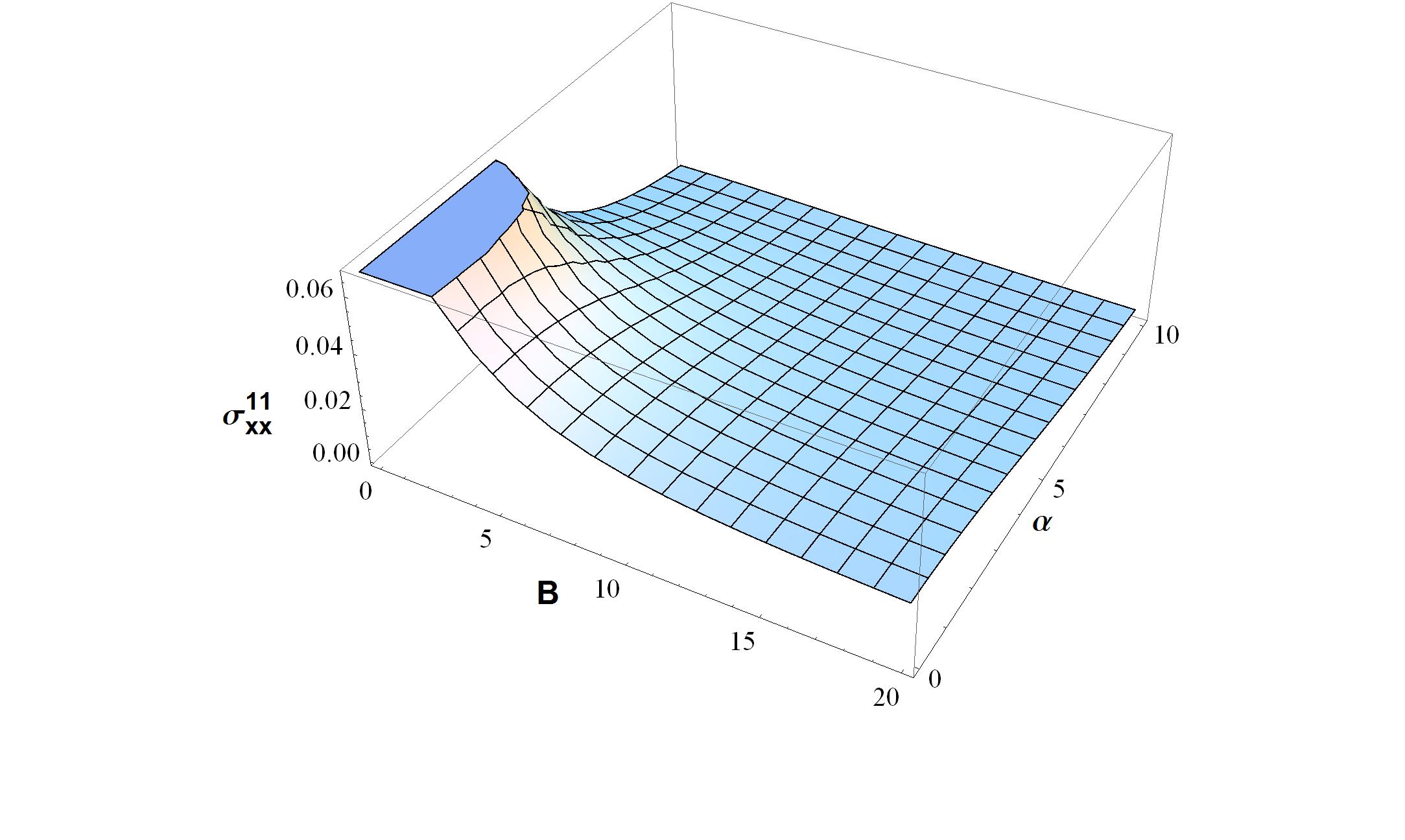}
\caption{Variation of $\sigma_{xx}^{11}$ with B and $\alpha$ for $z=1$(left) and $z=4/3$(right) }
\end{figure}
\begin{figure}[h!]
\centering
  \includegraphics[width=.45\textwidth]{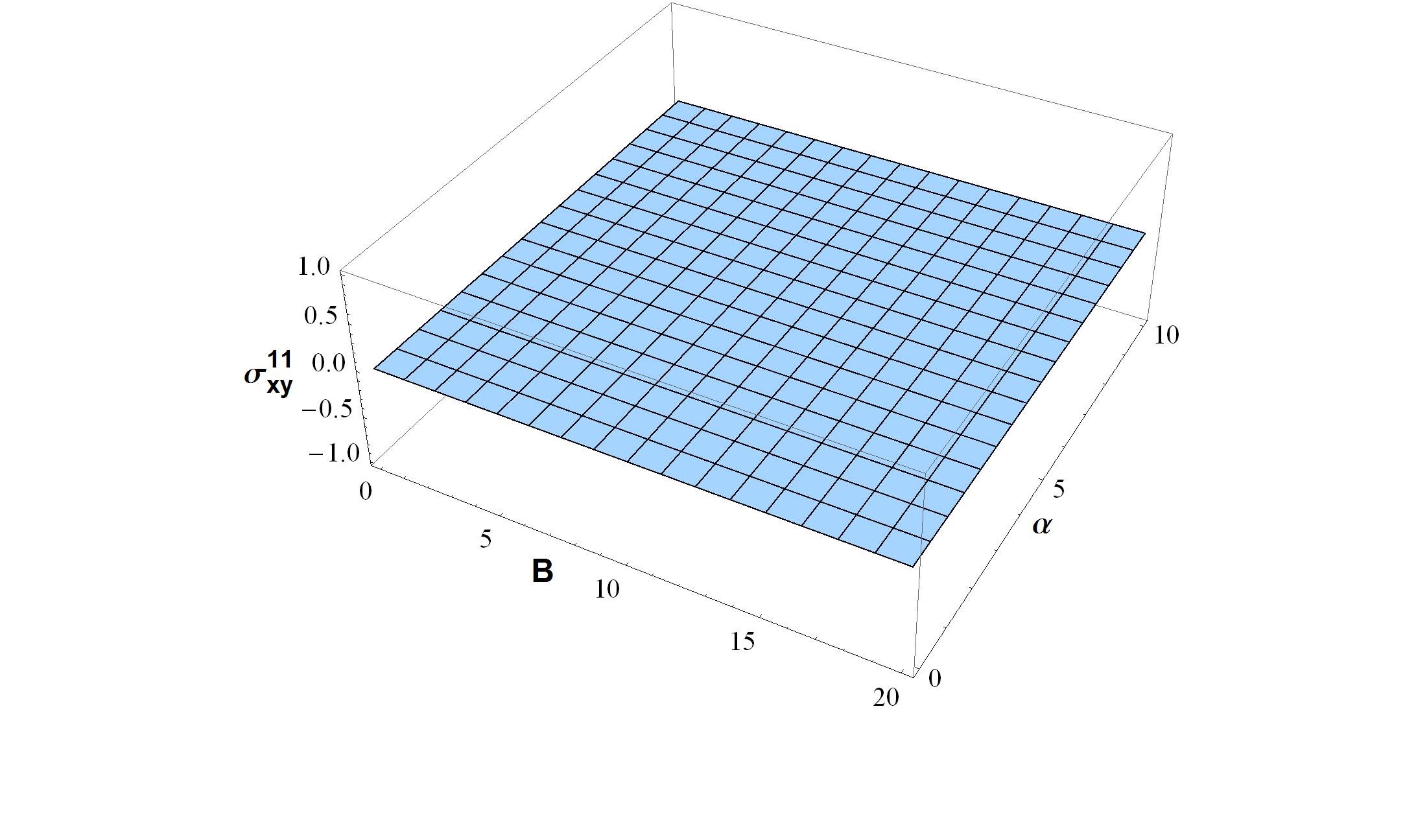}
  \includegraphics[width=.45\textwidth]{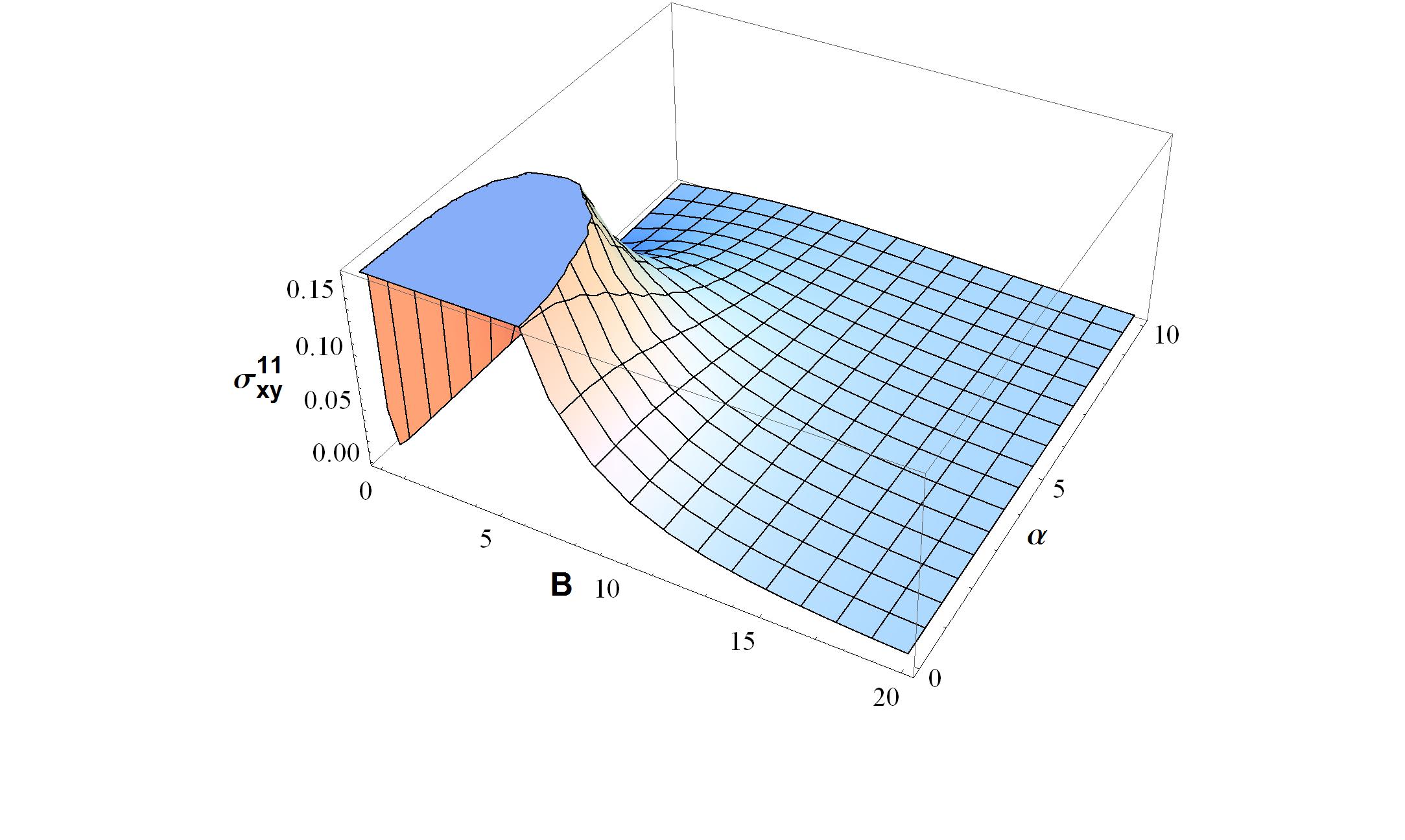}
\caption{Variation of $\sigma_{xy}^{11}$ with B and $\alpha$ for $z=1$(left) and $z=4/3$(right) }
\end{figure}
\begin{figure}[h!]
\centering
  \includegraphics[width=.45\textwidth]{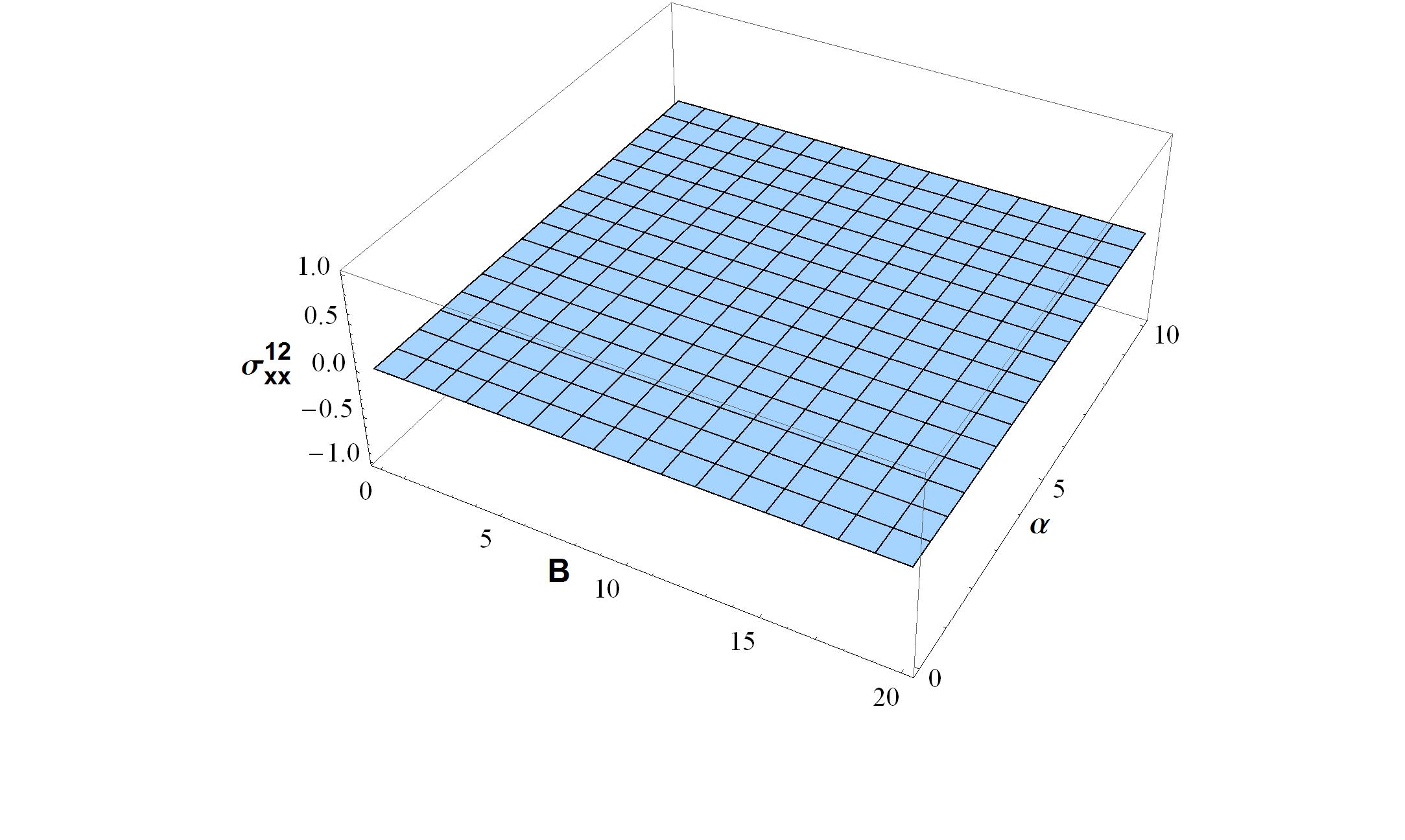}
  \includegraphics[width=.45\textwidth]{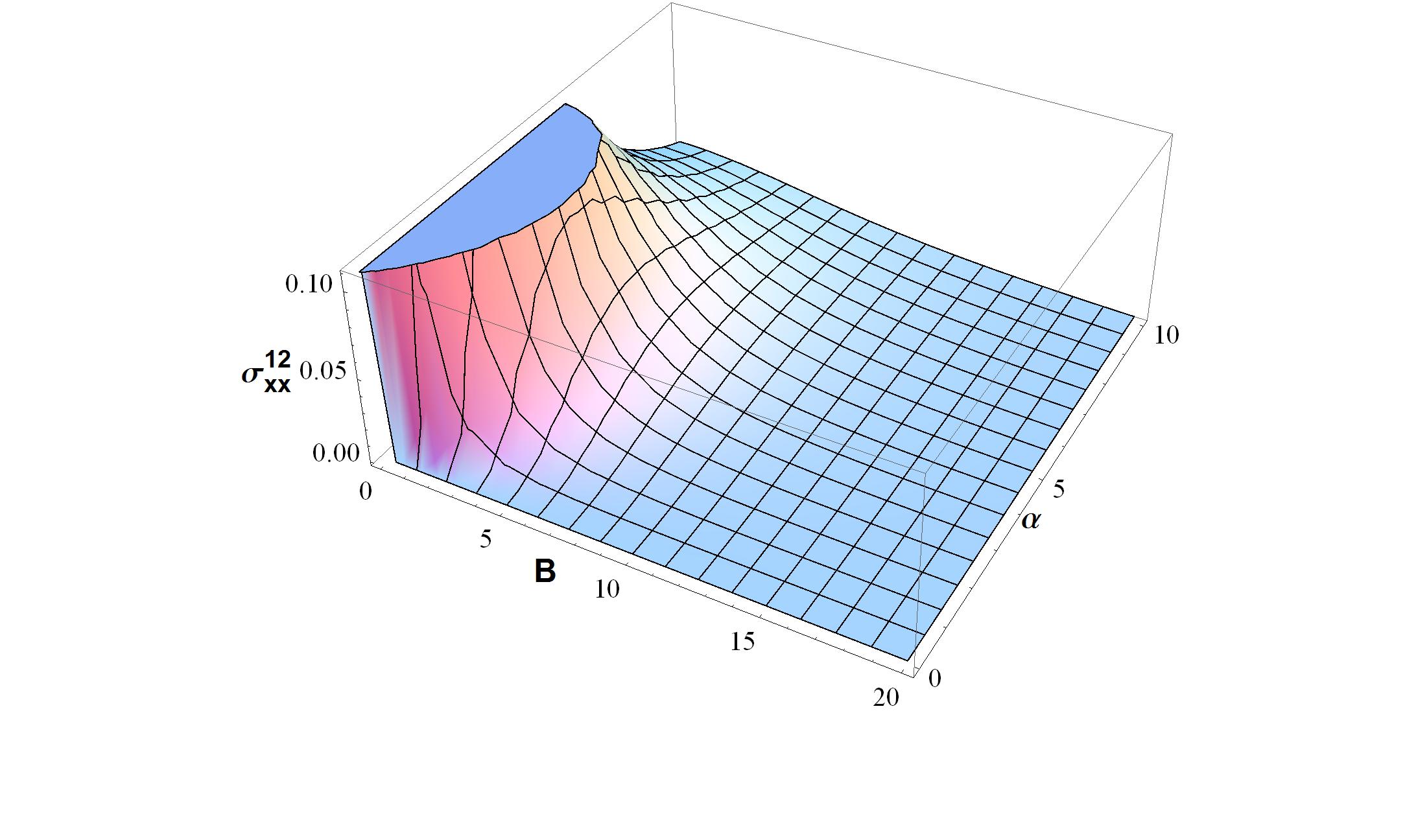}
\caption{Variation of $\sigma_{xx}^{12}$ with B and $\alpha$ for $z=1$(left) and $z=4/3$(right) }
\end{figure}
\begin{figure}[h!]
\centering
  \includegraphics[width=.45\textwidth]{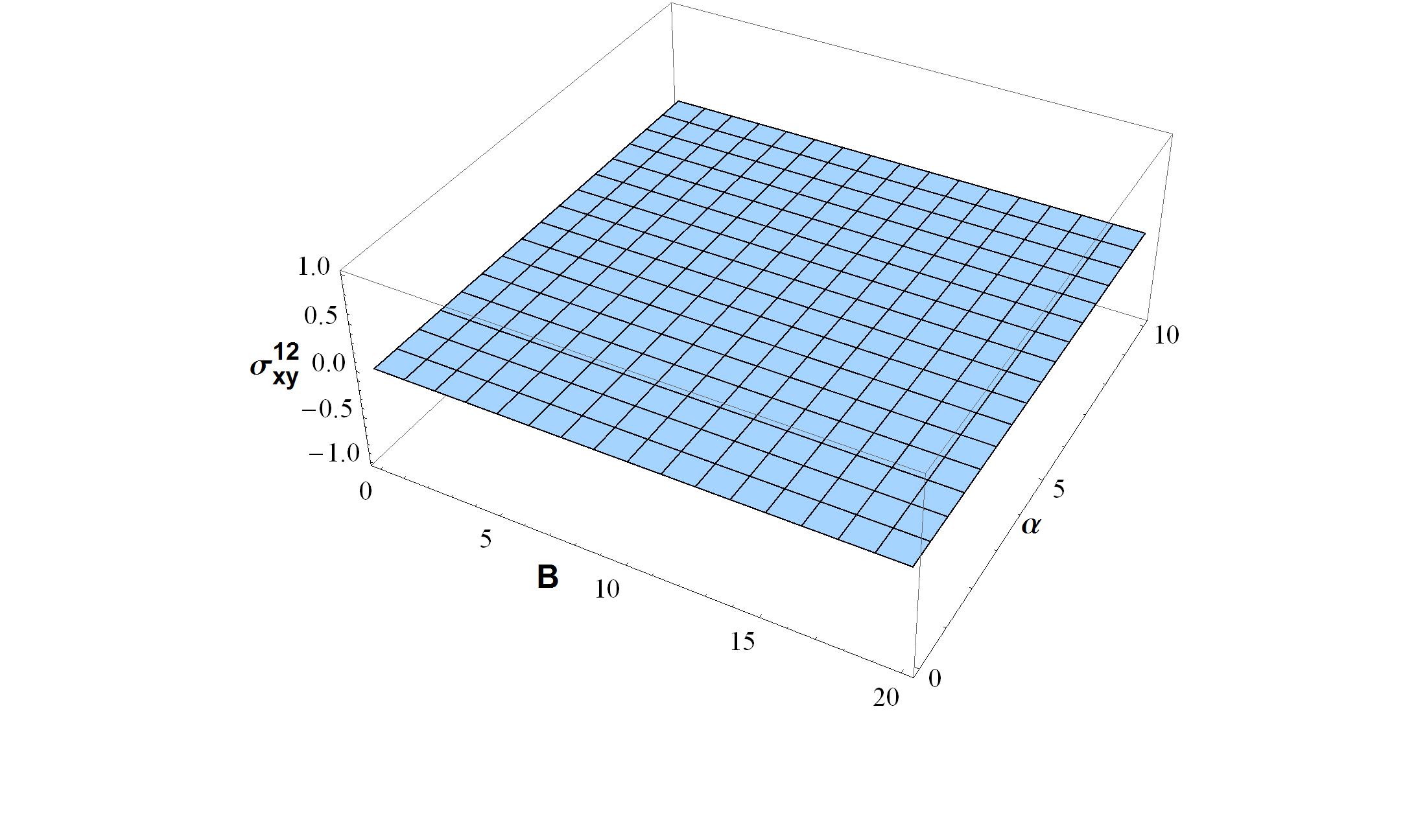}
  \includegraphics[width=.45\textwidth]{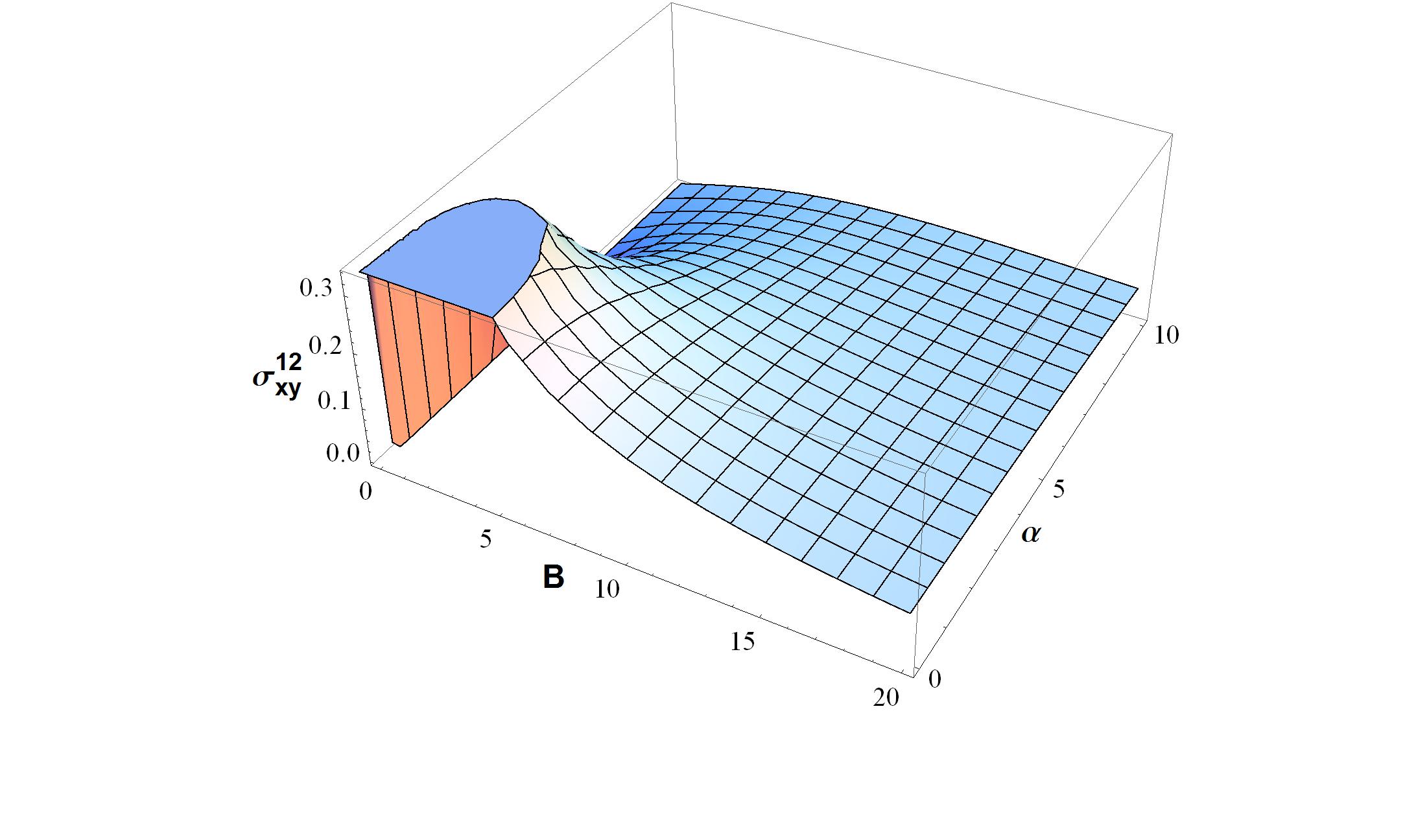}
\caption{Variation of $\sigma_{xy}^{12}$ with B and $\alpha$ for $z=1$(left) and $z=4/3$(right) }
\end{figure}

Keeping $z = 1$ reduces the geometry to RN-AdS and we observe trivial DC conductivity flow. However, for $z \neq 1$  non-trivial dependence of DC conductivity on magnetic field and momentum relaxation strength is seen. We also observe a discontinuity while considering $z=1$ ($\theta=0$) case, as $\sigma_{xx}^{11}=r_h^{-4}$. On the other hand for RN-AdS black hole the conductivity is given is a constant term given by, $\sigma_{xx}^{11}=1$.

Similarly, for the second gauge field we can evaluate the conductivity accordingly. The conserved current for the boundary theory is obtained using equation (\ref{eq:gauge1}),
\begin{equation}
J_{2i}=-r^{3z-1+\theta} f a'_{2i}-q_2h_{ti},
\end{equation}
On substituting equation (\ref{eq:f1}) we obtain the modified expression for $J_{2i}$ as,
\begin{equation}
J_{2i}=r_h^{-3+3z}i \omega a_{2i}+ \epsilon_{ij}iBr_h^{-3+3z}h_{tj}-q_2h_{ti}.
\end{equation}
Then the DC conductivity is evaluated concerning the second gauge field.  
\begin{eqnarray}
\sigma_{xx}^{22}&=&\sigma_{yy}^{22}=\frac{\alpha^2 r_h^{2z-4}[B^2+r_h^{-5z+7}(q_2^2r_h^{-z-1}+\alpha^2)]}{  B^2q_2^2r_h^{-2z-2}+(B^2r_h^{2z-4}+\alpha^2 r_h^{-3z+3})^2}, \label{eq:dc2}\\
\sigma_{xy}^{22}&=&-\sigma_{yx}^{22}=\frac{ q_2 B r_h^{4 z-8} \left[B^2+r_h^{-5 z+7} \left(q_2^2 r_h^{-z-1}+2 \alpha^2\right)\right]}{B^2q_2^2 r_h^{-2 z-2}+\left(B^2 r_h^{2 z-4}+\alpha ^2 r_h^{3-3 z}\right)^2}, \label{eq:dc3}
\end{eqnarray}
The above expressions are complicated to interpret. So we study the dependence of these conductivities on magnetic field and $\alpha$ using plots as shown in Fig.5 to  Fig.6. 
\begin{figure}[h!]
\centering
  \includegraphics[width=.45\textwidth]{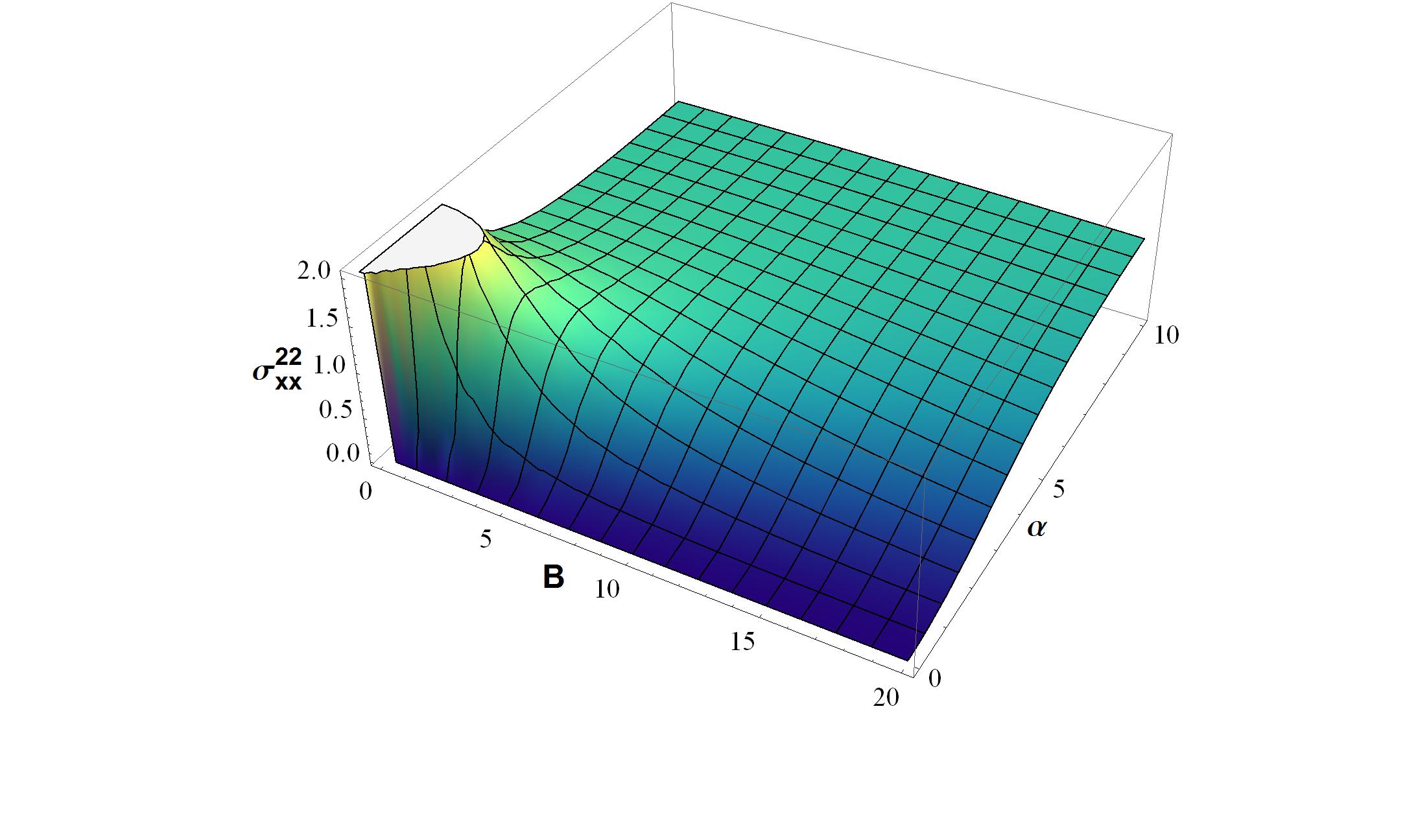}
  \includegraphics[width=.45\textwidth]{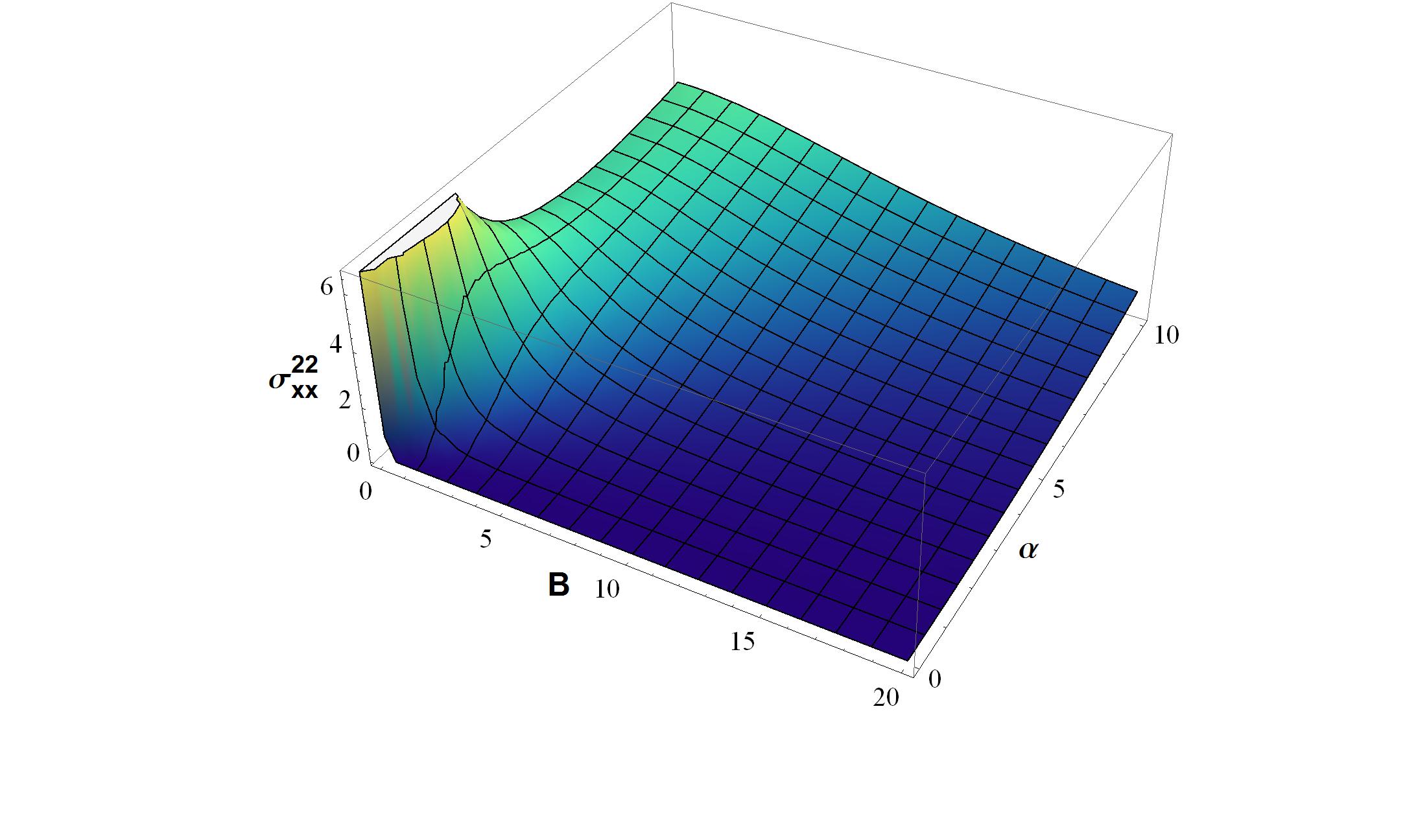}
\caption{Variation of $\sigma^{22}_{xx}$ with B and $\alpha$ for $z=1$(left) and $z=4/3$(right) }
\end{figure}
\begin{figure}[h!]
\centering
  \includegraphics[width=.45\textwidth]{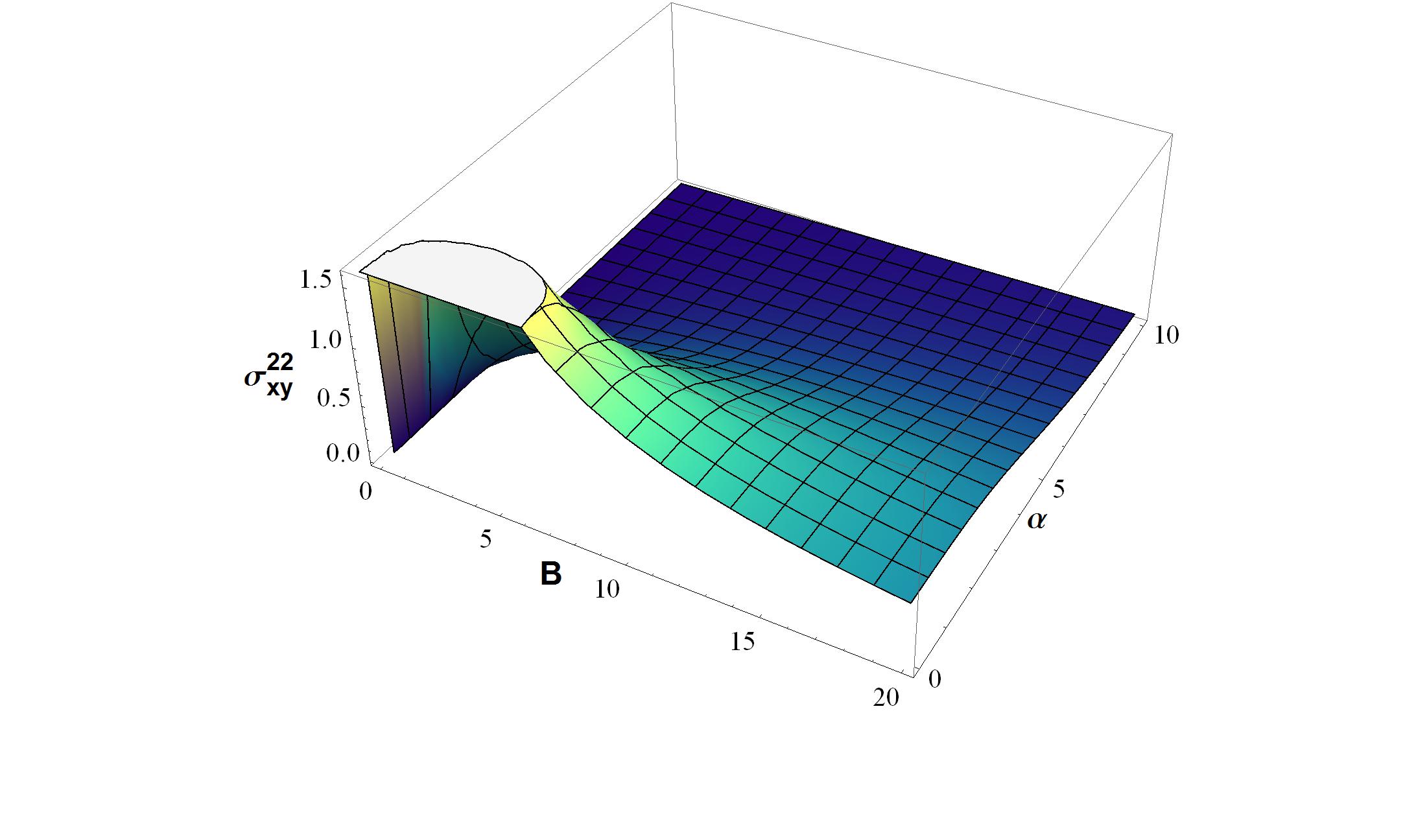}
  \includegraphics[width=.45\textwidth]{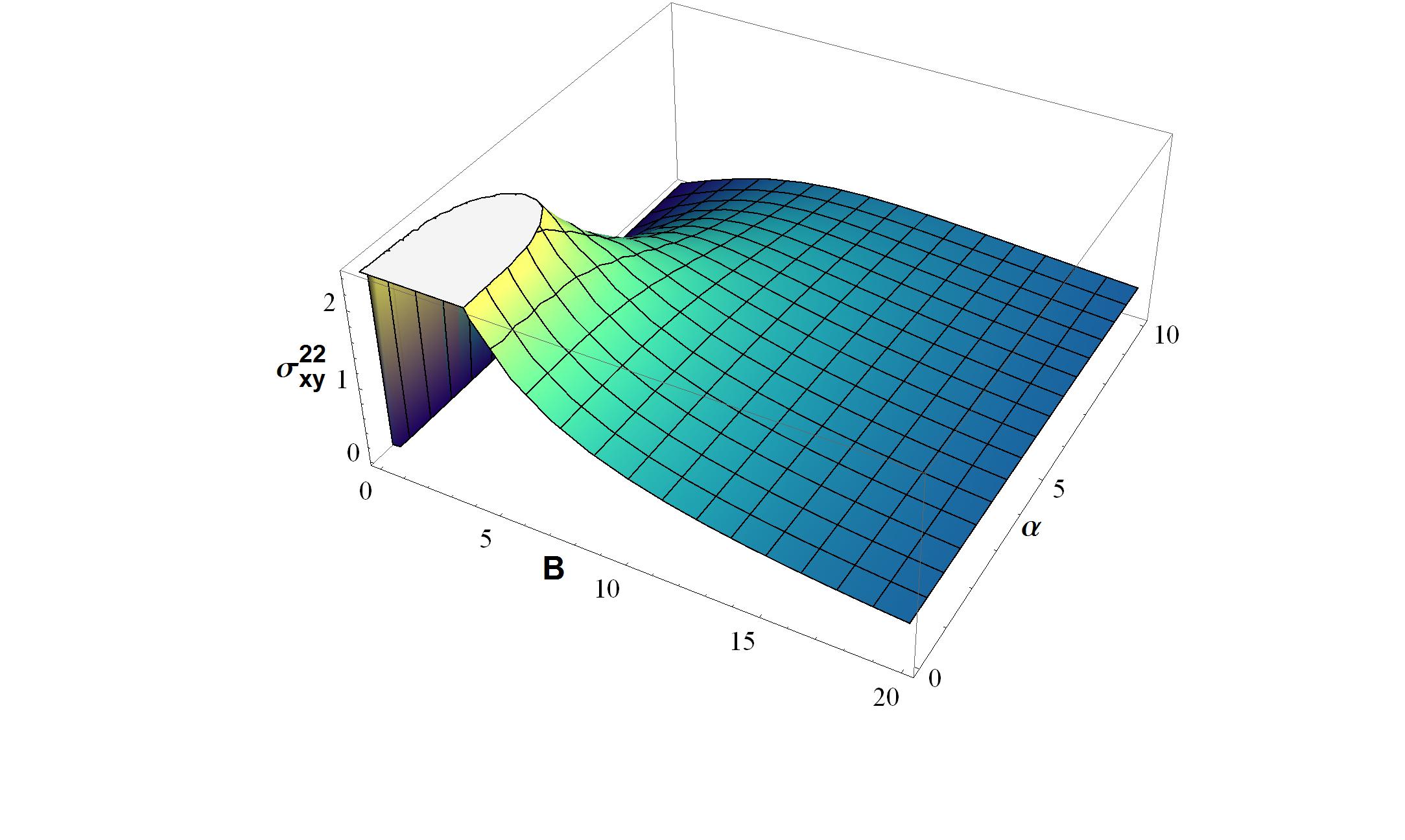}
\caption{Variation of $\sigma^{22}_{xy}$ with B and $\alpha$ for $z=1$(left) and $z=4/3$(right) }
\end{figure}

Also from the plots of different DC conductivities (for both the fields) we observe at fixed momentum relaxation strength conductivity $\sigma^{ij}_{xx}$ shows a monotonic dependence on the external magnetic field whereas in $\sigma^{ij}_{xy}$ this behavior is not seen. \\
Let us consider the limit $B\rightarrow 0$. 
\begin{equation}
\sigma^{22}_{xx}= r_h^{3z-3}+r_h^{2z-4}\frac{q_2^2}{\alpha^2},\qquad \sigma^{22}_{xy}=0\label{eq:d}
\end{equation}
From the above expression it is noted that electric conductivity obeys inverse Matthiessen's rule given by,
\begin{equation}
\sigma_{DC}=\sigma_{Q}+\sigma_{D}
\end{equation}
where $\sigma_{Q}$ is the charge conjugation symmetric part and $\sigma_{D}$ is the momentum dissipation part. 

The temperature dependence of conductivity is governed by equation ($\ref{eq:T}$)  and $T \sim r_h^z$. Here we observe the following scaling in the DC conductivity,
\begin{itemize}
\item[i] For $z=1$, \quad $\sigma^{22}_{xx} \sim 1+\frac{q_2^2}{\alpha^2 T^2}$
\item[ii] For $z=4/3$,\quad $\sigma^{22}_{xx} \sim T^{3/4}+\frac{q_2^2}{T\alpha^2}$
\item[ii] For $z \rightarrow 2$, \quad $\sigma^{22}_{xx} \sim T^{3/2}+\frac{q_2^2}{\alpha^2}$
\end{itemize}
In our holographic model, we are able to capture the low temperature behavior of the DC conductivity obeying Fermi-Liquid law ( $\sigma_{DC} \sim \frac{1}{T^2}$ ) for $z=1$ along with a constant term.  This behavior changes to unconventional metallic behavior ($ \sigma_{DC} \sim \frac{1}{T}$) as we increase the Lifshitz scaling to $z=4/3$ and becomes constant in the limiting case $z\rightarrow 2$. Thus, there is non-trivial dependence of conductivity on temperature for hyperscaling range ($1<z<2$).  

\subsection{Halls Angle}
We can obtain the expression for Hall angle using equations (\ref{eq:dc2}) and (\ref{eq:dc3})\cite{Kim:2015wba,Ge:2016sel,Amoretti:2017xto}.
Thus, 
\begin{equation}
 \tan{ \theta_H}=\frac{\sigma^{22}_{xy}}{\sigma^{22}_{xx}}
\end{equation}
\begin{equation}
\theta_H=\frac{B q_2 r_h^{2 z-4} \left[B^2+ r_h^{7-5 z} \left(2 \alpha ^2+q_2^2 r_h^{-z-1}\right)\right]}{\alpha ^2 \left[B^2+r_h^{7-5 z} \left(\alpha ^2+q_2^2 r_h^{-z-1}\right)\right]} \label{eq:ha1}
\end{equation}
From the above expression it is observed that, $\theta_H \propto \frac{B q_2  r_h^{2z-4}}{\alpha^2}$ as the terms in the bracket is consider as a geometric quantity \cite{Blake:2014yla}. 

Comparing the result with that of DC conductivity Hall angle consists of only dissipation part $(\sigma_{D})$, unlike DC conductivity which is the combination of two different terms (shown in equation (\ref{eq:d})). This is responsible for the presence of different scaling in strange metals\cite{Andri:1997}. We plot the dependence of Hall angle on the magnetic field applied and the strength of momentum relaxation in Fig.7 for fixed $q_2$.  
\begin{figure}[h!]
\centering
  \includegraphics[width=.45\textwidth]{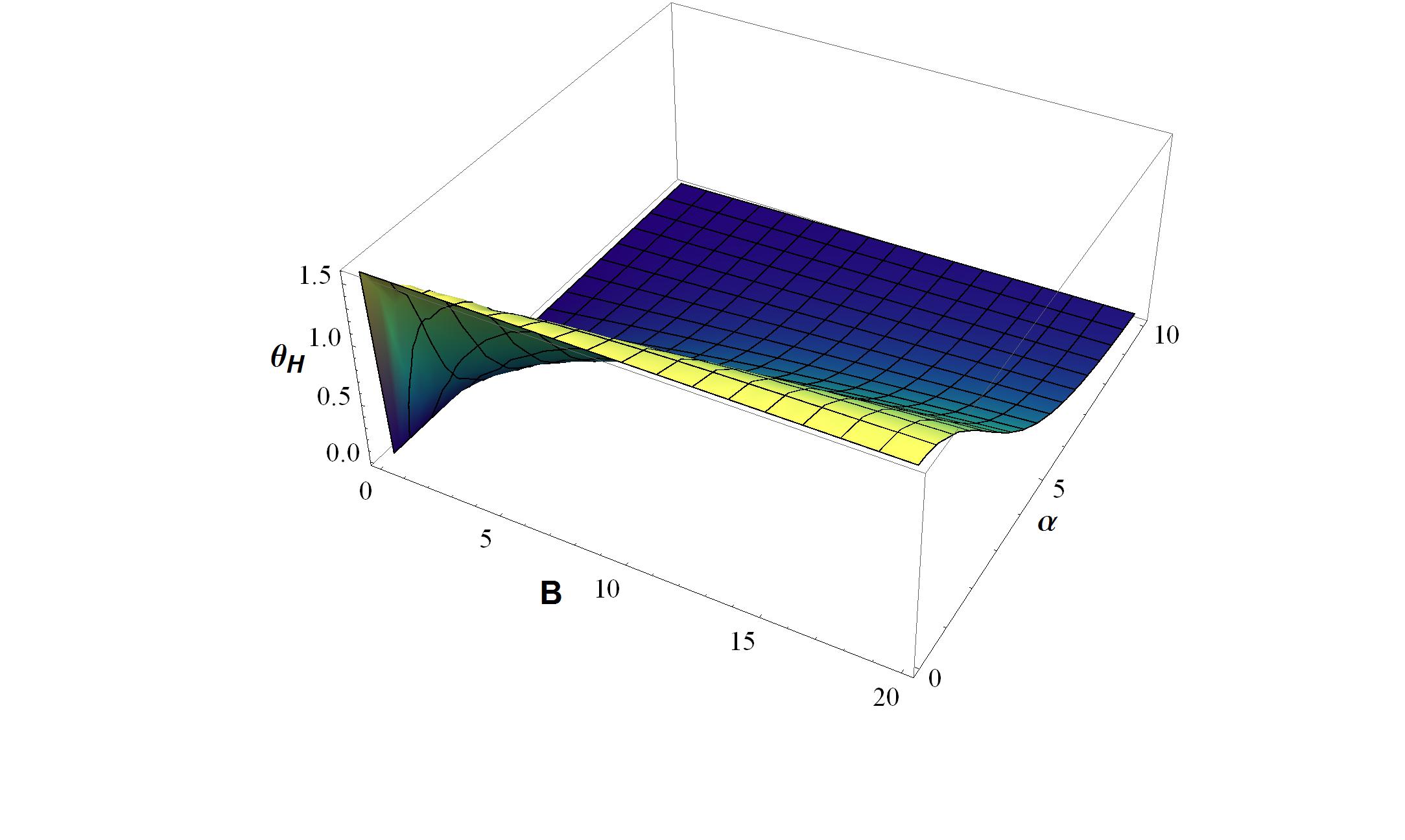}
  \includegraphics[width=.45\textwidth]{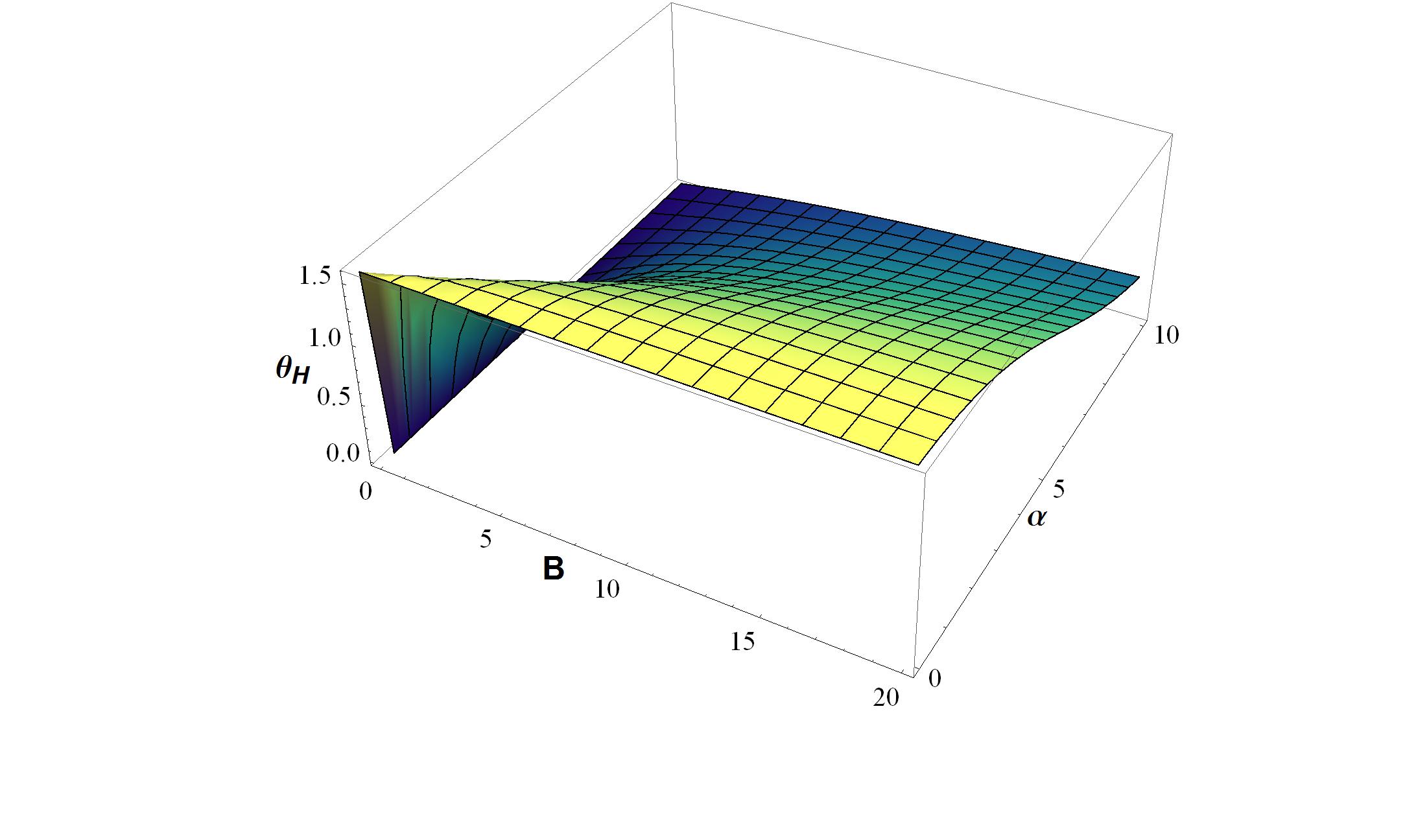}
\caption{Variation of  $\theta_H$  with B and $\alpha$ for $z=1$(left) and $z=4/3$(right) }
\end{figure}
Let us consider the temperature dependence of Hall angle,
\begin{itemize}
\item[i] For $z=1$, \quad $\theta_H \sim 1/T^2$
\item[ii] For $z=4/3$,\quad $\theta_H \sim 1/T$
\item[ii] For $z \rightarrow 2$, \quad $\theta_H \sim 1/T^0$
\end{itemize}
For $z=1$, we observe the temperature dependence is same as measured in cuprates \cite{Hussey:2008}. However the behavior changes to $\theta_H \sim 1/T$ with the non-trivial scaling $z \neq 4/3$. Further $\theta_H$ reduces to a constant for $z \rightarrow 2$. We show the temperature and magnetic field dependence on the Hall angle from plots given in Fig.8 and Fig.9.

\begin{figure}[h!]
\centering
  \includegraphics[width=.38\textwidth]{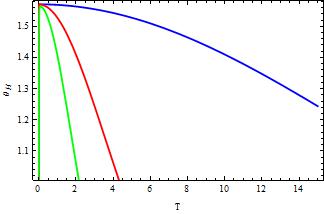}
\hspace{1cm}
  \includegraphics[width=.38\textwidth]{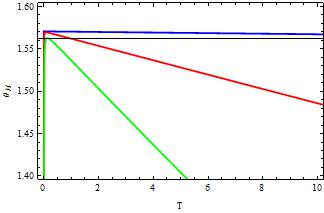}
\caption{$\theta$ vs T at $\alpha$= 0.1(blue),0.5(red),1(green) for $z=1$(left) and $z=4/3$(right)}
\end{figure}
\begin{figure}[h!]
\centering{
  \includegraphics[width=.38\textwidth]{ha4.jpg}
\hspace{1cm}
  \includegraphics[width=.38\textwidth]{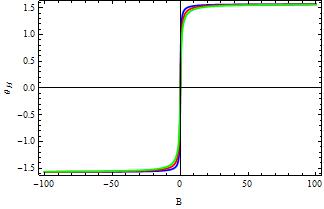}
\caption{$\theta$ vs B at $T=$ 0.5(blue),1(red),1.5(green) for $z=1$(left) and $z=4/3$(right) }}
\end{figure}
\subsection{Thermoelectric conductivity}
Next, we evaluate thermoelectric conductivity using heat current expression. 
\begin {eqnarray}
Q_i=-4\pi T g_{xx}h_{ti}, \quad {\text{and}} \quad \alpha_{ij} =\frac{\partial Q_i}{T \partial E_j}
\end{eqnarray}
We get the following expression for thermoelectric conductivity depending on external magnetic field and dynamical exponent for our model as shown in Fig. 10 and Fig. 11.
\begin{eqnarray}
\alpha^{22}_{xx}=\alpha^{22}_{yy}=\frac{4 \pi r_h^{-3z+3}q_2 \alpha^2}{(B^2r_h^{2z-4}+\alpha^2 r_h^{-3z+3})^2+B^2q_2^2r_h^{-2z-2}},
\end{eqnarray}
\begin{eqnarray}
\alpha^{22}_{xy}=-\alpha^{22}_{yx}=\frac{4 \pi B r_h^{z+1}(B^2 r_h^{4z-8}+\alpha^2 r_h^{-z-1}+q_2^2 r_h^{-2-2z})}{(B^2r_h^{2z-4}+\alpha^2 r_h^{-3z+3})^2+B^2q_2^2r_h^{-2z-2}}.
\end{eqnarray}

In the limiting case $B\rightarrow 0$, we obtain,
 \begin{equation}
\alpha^{22}_{xx}=\frac{4 \pi q_2 r_h^{3z-3}}{\alpha^2}, \qquad \alpha^{22}_{xy}=0 \label{eq:c1}
\end{equation}
Thus, the thermoelectric conductivity also shows non-trivial dependence on hyperscaling and unconventional temperature dependence. 

\begin{figure}[h!]
\centering
  \includegraphics[width=.45\textwidth]{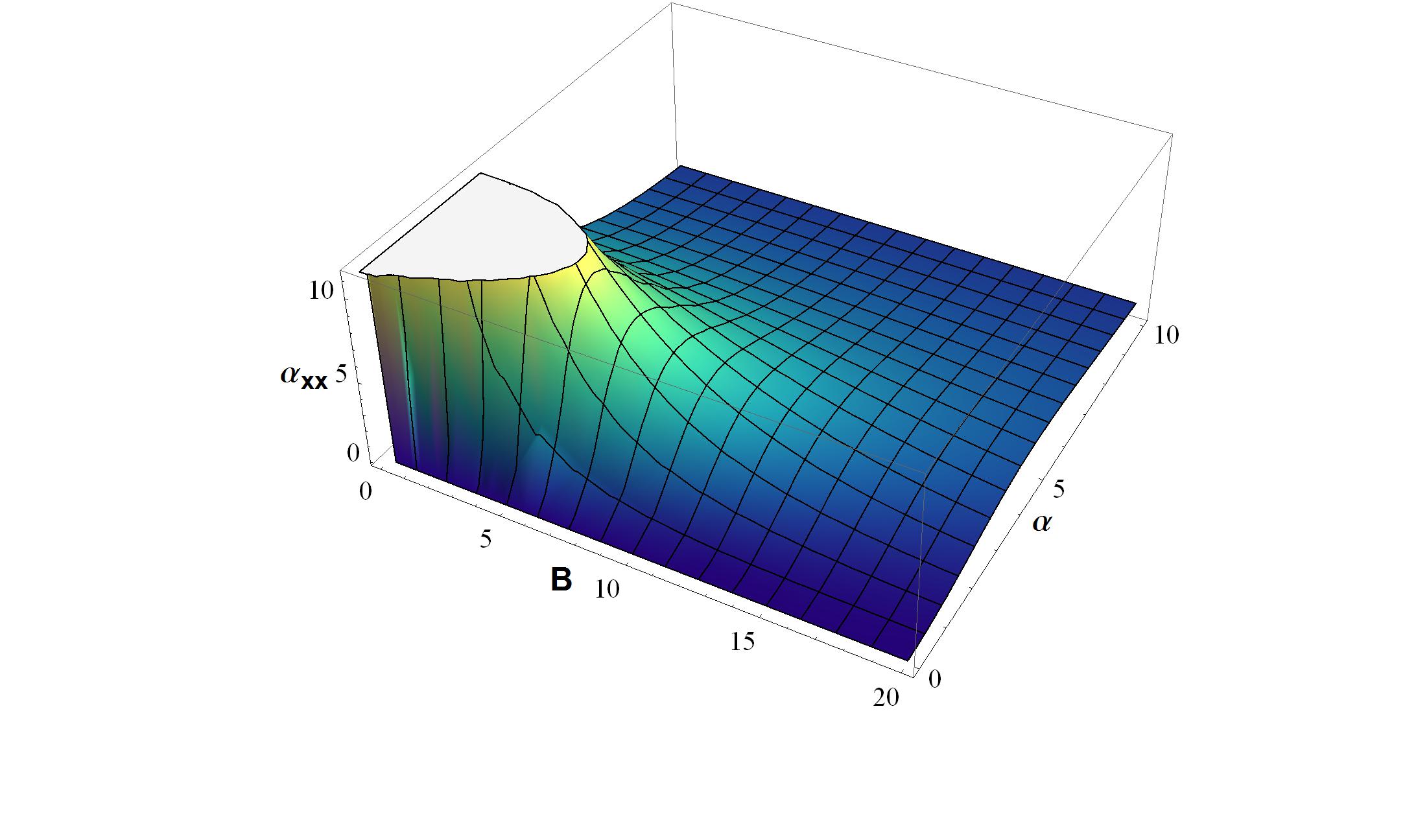}
  \includegraphics[width=.45\textwidth]{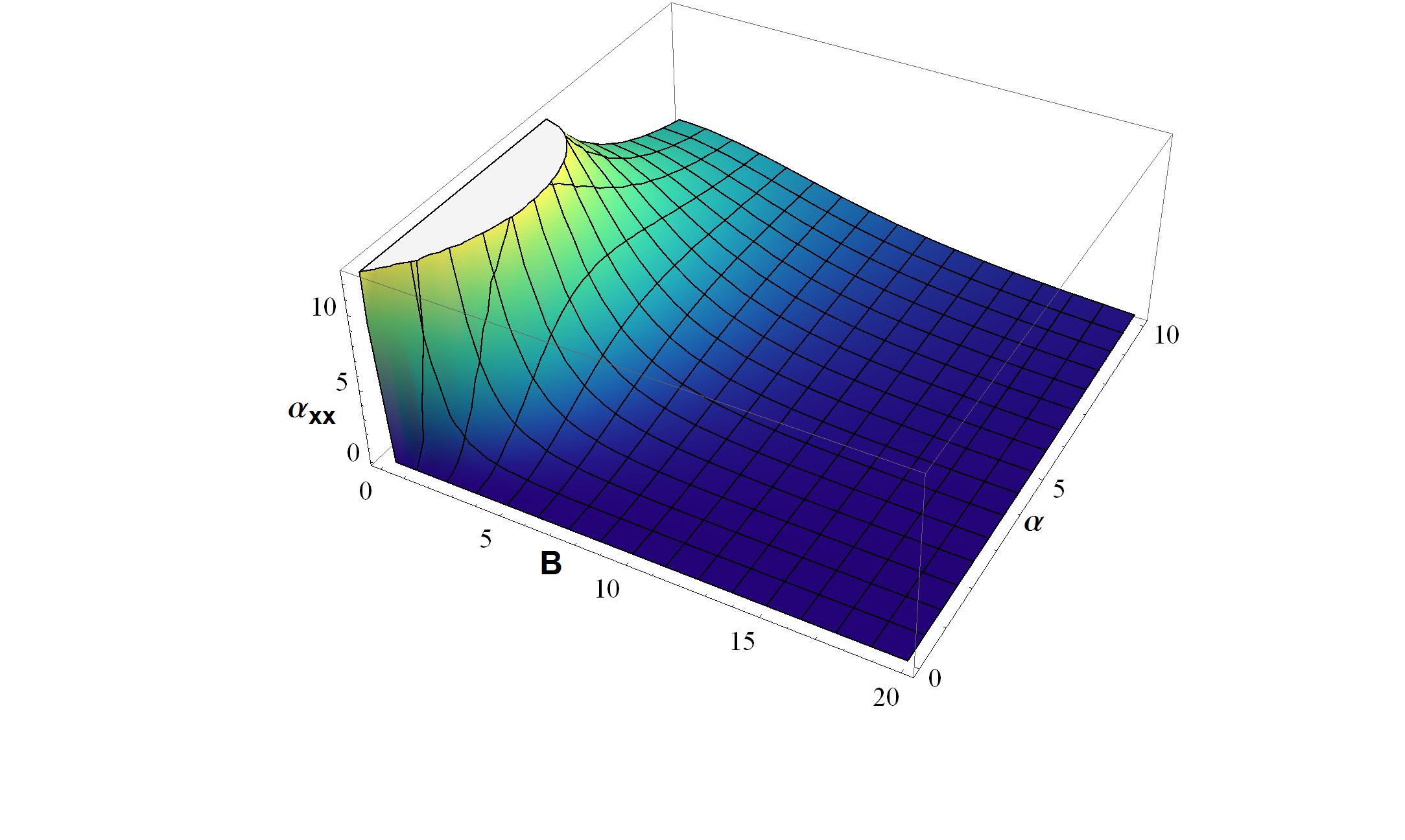}
\caption{Variation of  $\alpha^{22}_{xx}$ with B and $\alpha$ for $z=1$(left) and $z=4/3$(right) }
\end{figure}
\begin{figure}[h!]
\centering
  \includegraphics[width=.45\textwidth]{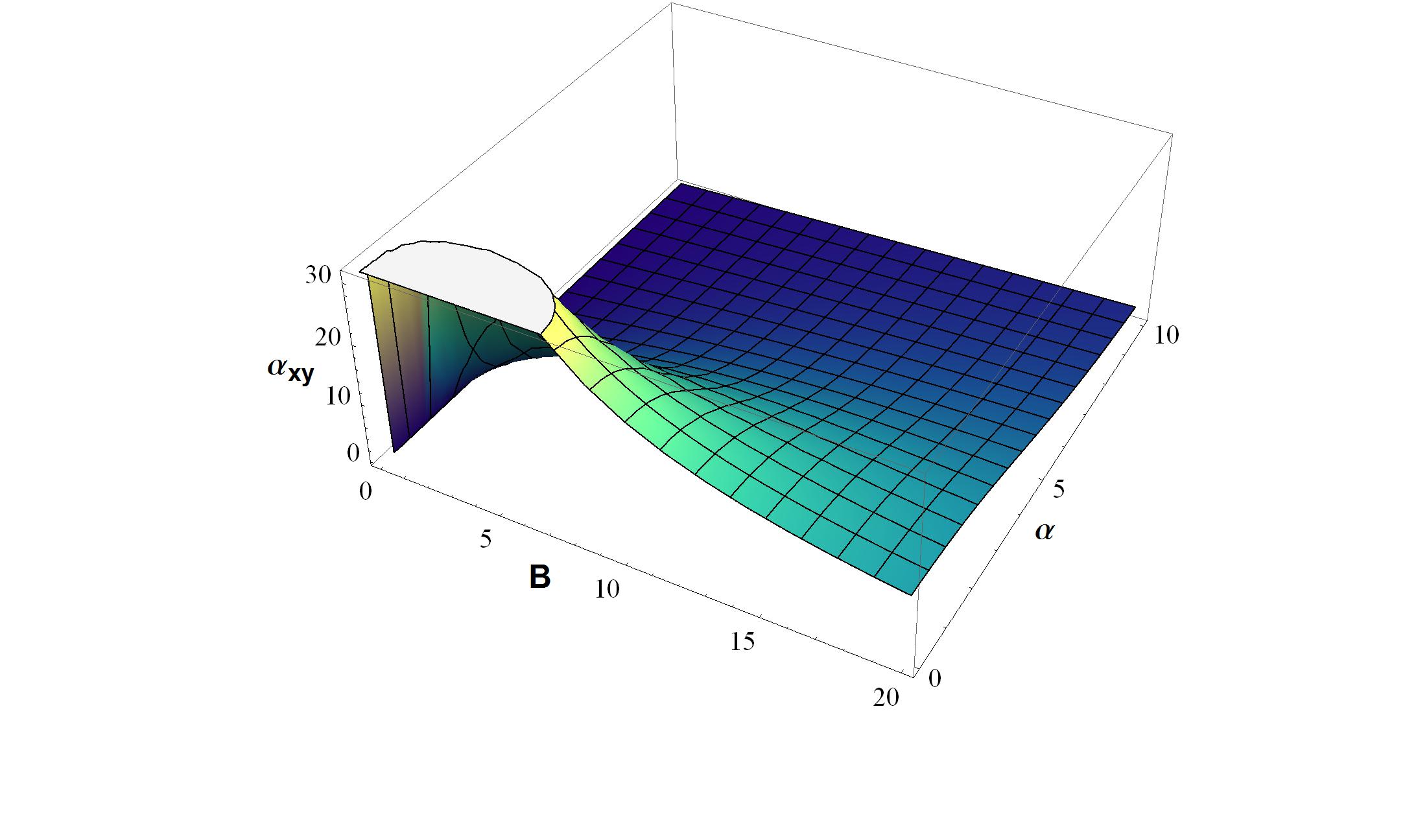}
  \includegraphics[width=.45\textwidth]{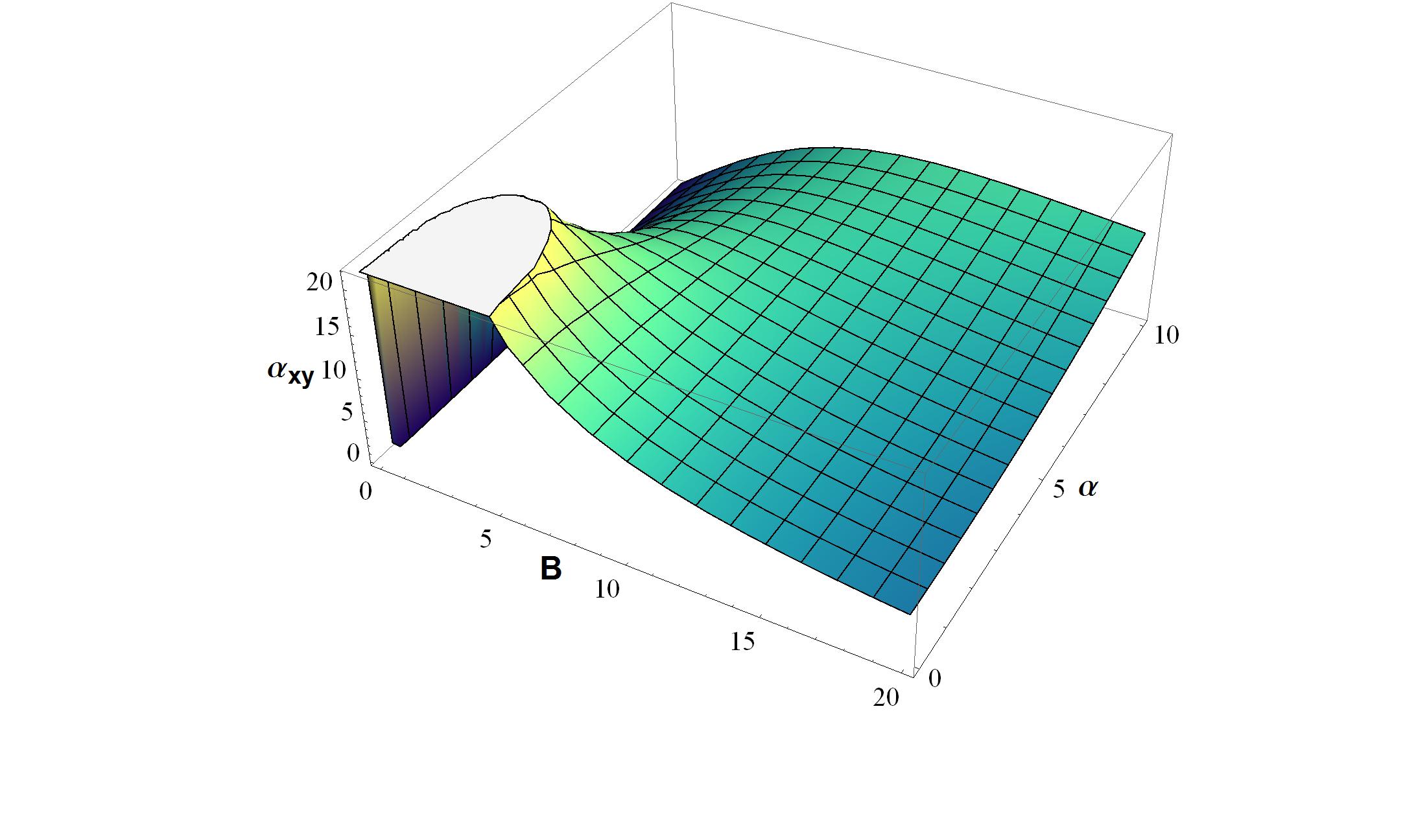}
\caption{Variation of  $\alpha^{22}_{xy}$ with B and $\alpha$ for $z=1$(left) and $z=4/3$(right) }
\end{figure}
\subsection{Seebeck Coefficient} 
The generation of transverse electric field in the system is given by the thermoelectric power (Seebeck coefficient). Using the results of the conductivity we obtain,   
\begin{equation}
S=\frac{\alpha^{22}_{xx}}{\sigma^{22}_{xx}}=\frac{4 \pi r_h^{-5z+7}q_2}{[B^2 + r_h ^{-5z+7}(q_2^2 r_h^{-z-1}+\alpha^2)]} \cdot
\end{equation}

\begin{figure}[h!]
\centering
  \includegraphics[width=.45\textwidth]{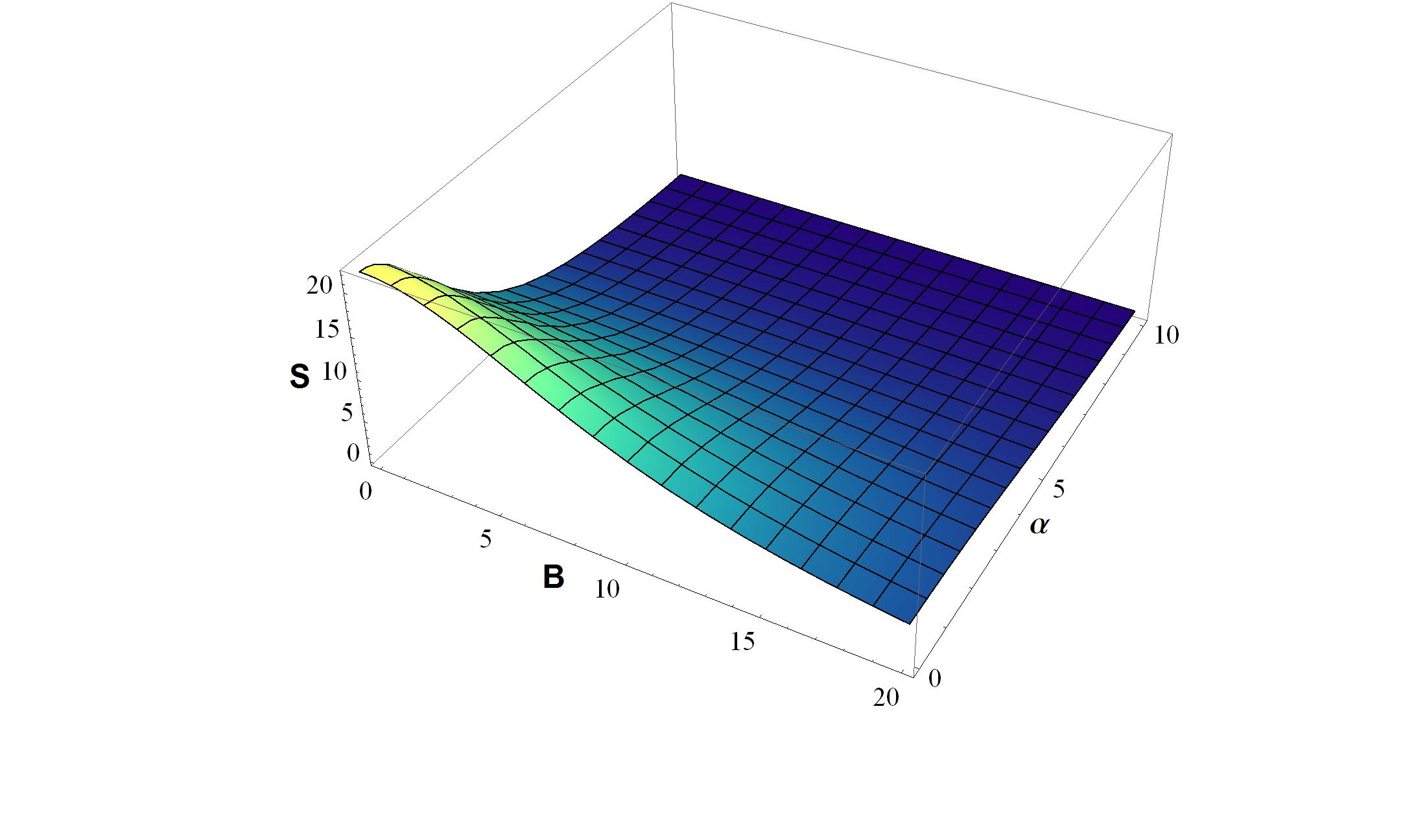}
  \includegraphics[width=.45\textwidth]{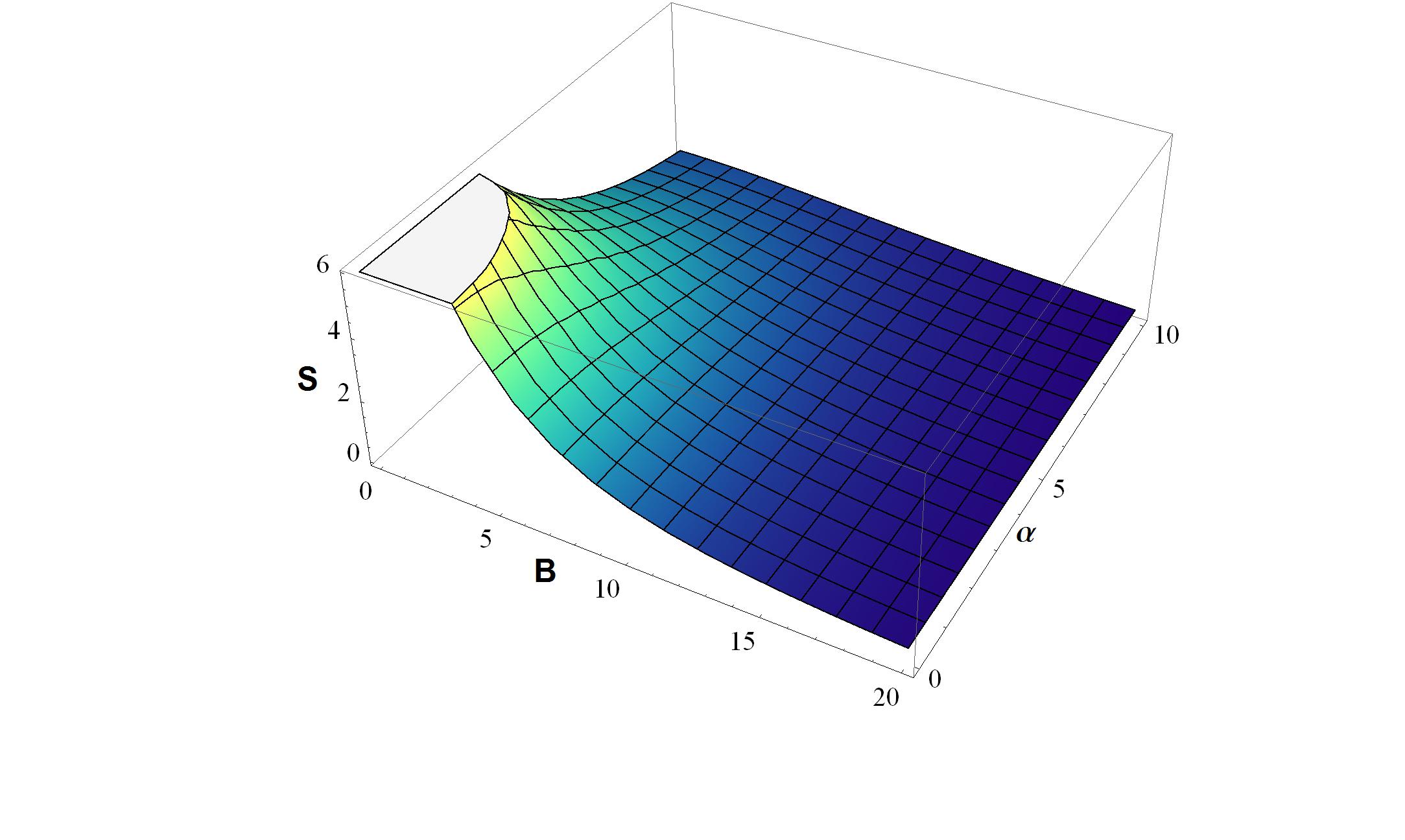}
\caption{Variation of S with B and $\alpha$ for $z=1$(left) and $z=4/3$(right) }
\end{figure}

\begin{figure}[h!]
\centering
  \includegraphics[width=.38\textwidth]{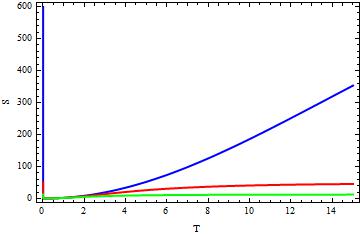}
\hspace{1cm}
  \includegraphics[width=.38\textwidth]{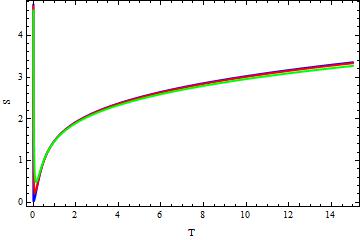}
\caption{S vs T at $\alpha$=0.5(blue),1(red),1.5(green) for $z=1$(left) and $z=4/3$(right) }
\end{figure}
\begin{figure}[h!]
\centering
  {\includegraphics[width=.38\textwidth]{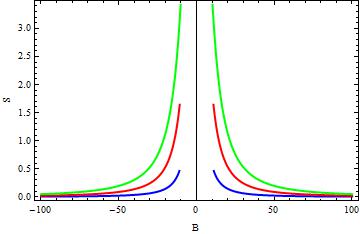}
\hspace{1cm}
  \includegraphics[width=.38\textwidth]{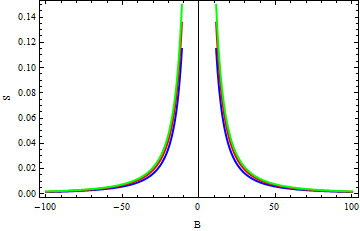}}
\caption{S vs B at $T$=0.5(blue),1(red),1.5(green) for $z=1$(left) and $z=4/3$(right) }
\end{figure}
The variation of Seebeck coefficient with applied magnetic field and momentum relaxation strength is shown in Fig. 12. The result from experiments suggest that at high temperature Seebeck coefficient remains constant. The dependence of Seebeck coefficient on model parameters is shown in Fig.13 and Fig.14. We observe at different temperature the behavior of  Seebeck coefficients does not change appreciably for non-trivial scaling. The temperature scaling of the coefficient is still unclear from experimental results\cite{Kim:2004}.
\subsection{Thermal conductivity}
Using the results for thermoelectric and DC conductivity, we can also obtain the thermal conductivity \cite{Tian:2017vfk}.
The thermal conductivity for non-zero magnetic field can be obtained using the relation given below,
\begin{equation}
\begin{pmatrix} <J_{i}> \\ <Q_{i}> \end{pmatrix} = \begin{pmatrix}\sigma_{ij}& \alpha_{ij}T\\ \bar{\alpha}_{ii}T & \bar{\kappa}_{ij}T \end{pmatrix} \begin{pmatrix}E_j \\ -(\nabla_j T)/T\end{pmatrix}
\end{equation}
Considering the thermal current $Q_x=0$  and $E_y=0$ we obtain the expression for the thermal conductivity using,
\begin{equation}
\bar{\kappa}^{22}_{xx}=\frac{T (\alpha^{22}_{xx})^2}{\sigma^{22}_{xx}-\sigma^{22}_{xx}(0)}, \quad {\text{and}}  \quad \bar{\kappa}^{22}_{xy}=\frac{T \alpha^{22}_{xx} \alpha^{22}_{xy}}{\sigma^{22}_{xx}}
\end{equation}
where $\sigma^{22}_{xx}(0)$ is the electric conductivity for $Q_x=0$ (vanishing heat currents).
Thus, we obtain, 
\begin{equation}
\bar{\kappa}^{22}_{xx}=\bar{\kappa}^{22}_{yy}=\frac{16 \pi^2 r_h^{5 + 4 z} T (B^2 r_h^{5 z}+ r_h^7 \alpha^2)}{
B^4 r_h^{10 z} + r_h^{14} \alpha^4 + 
 B^2 r_h^{6 + 4 z} (q_2^2 + 2 r_h^{1 + z}\alpha^2)}
\end{equation}
\begin{equation}
\bar{\kappa}^{22}_{xy}=-\bar{\kappa}^{22}_{yx}=\frac{16 \pi ^2 T B q_2 r_h^{8+6z}}{B^4 r_h^{10z}+r_h^{14}\alpha^2 +B^2 r_h^{6+4z}(q_2^2+2 r_h^{1+z}\alpha^2)}
\end{equation}
Variation of thermal conductivity with different model parameters is shown in Fig.15 and Fig.16
\begin{figure}[h!]
\centering
  \includegraphics[width=.45\textwidth]{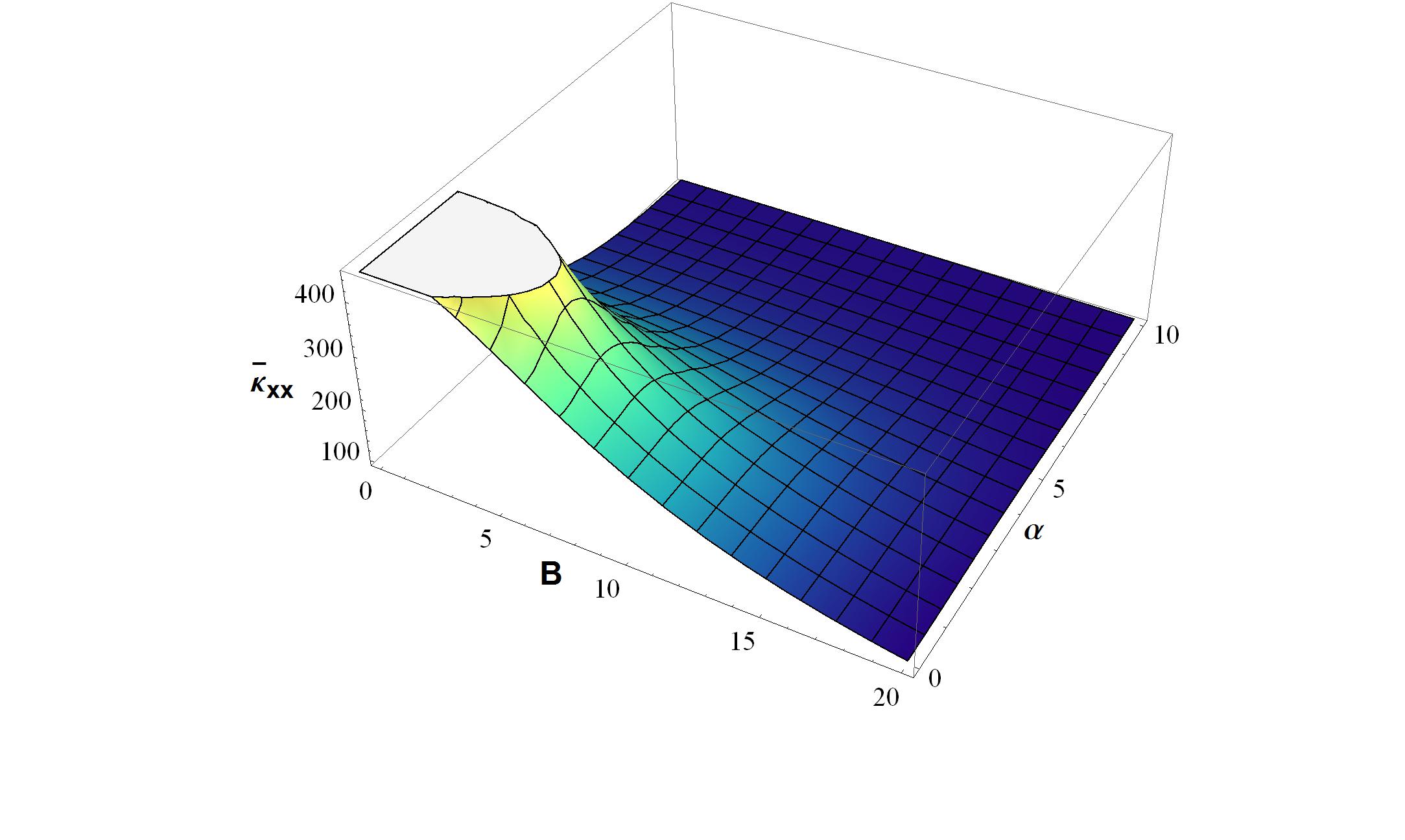}
  \includegraphics[width=.45\textwidth]{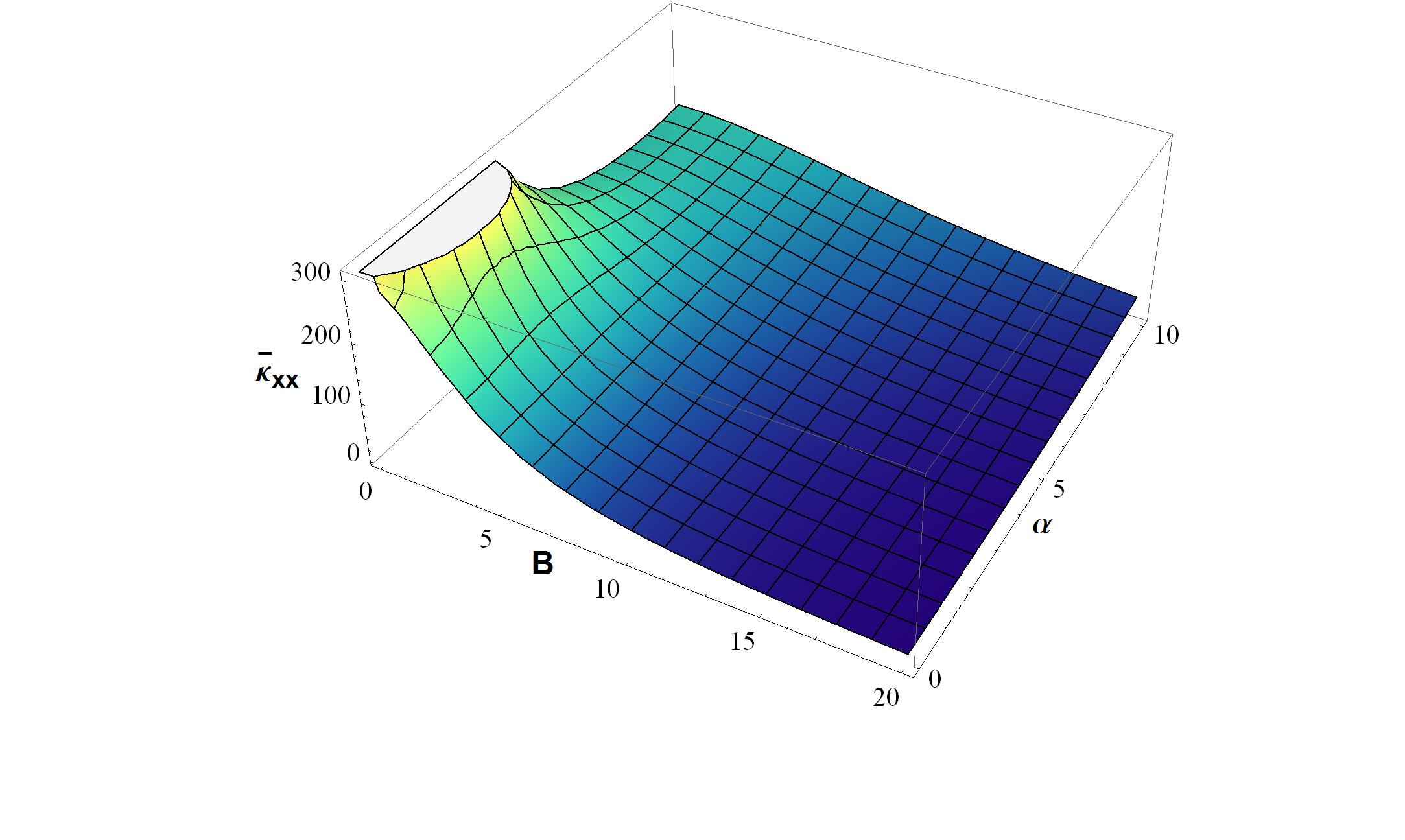}
\caption{Variation of $\bar{\kappa}^{22}_{xx}$  with B and $\alpha$ for $z=1$(left) and $z=4/3$(right) }
\end{figure}
which also shows non-trivial dependence on hyperscaling.  

Let us discuss the limiting case $B\rightarrow 0$,
\begin{equation}
\kappa^{22}_{xx}=\frac{16 \pi ^2 T r_h^{4 z-2}}{\alpha ^2}, \qquad \kappa^{22}_{xy}=0,
\end{equation}

\begin{figure}[h!]
\centering
  \includegraphics[width=.45\textwidth]{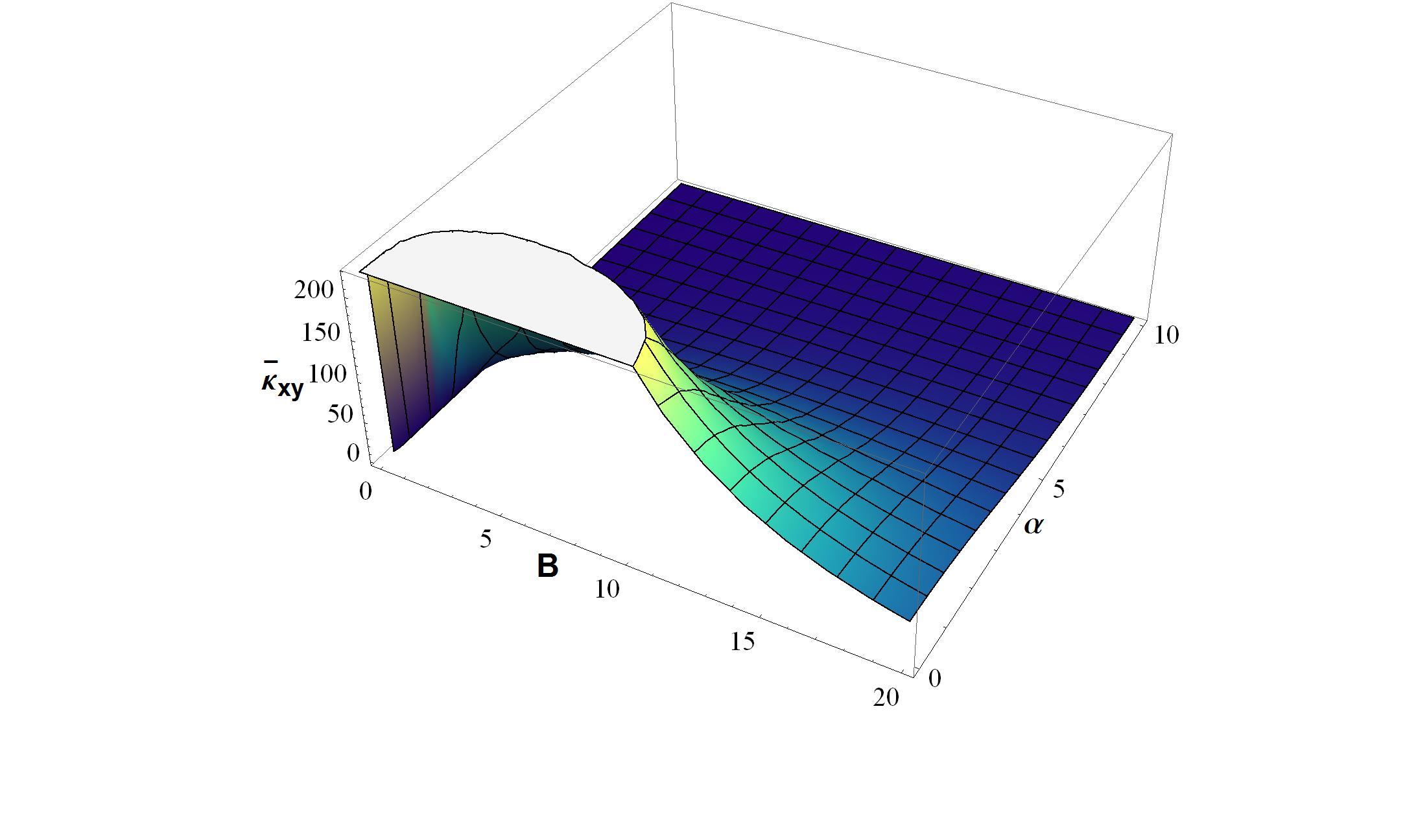}
  \includegraphics[width=.45\textwidth]{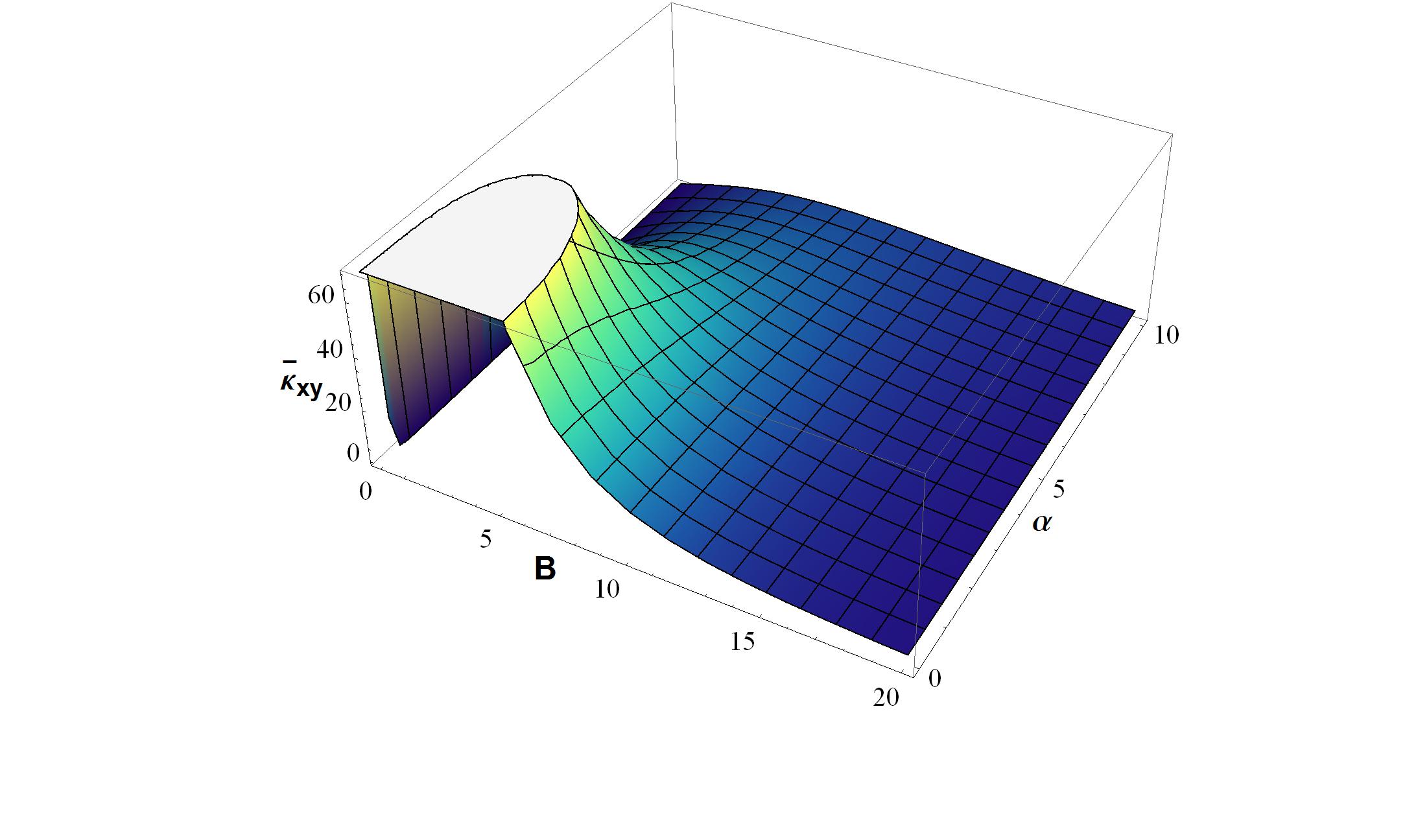}
\caption{Variation of $\bar{\kappa}^{22}_{xy}$  with B and $\alpha$ for $z=1$(left) and $z=4/3$(right) }
\end{figure}
\subsection{Lorenz ratio}  
To complete the discussion, we check the expression for Lorenz ratio. 
First we obtain the Hall Lorentz ratio using \cite{Ge:2016sel},
\begin{equation}\label{eq:L1}
L=\frac{ \bar{\kappa}_{xy}}{T \sigma_{xy}}=\frac{16 \pi ^2 r_h^{2 z+8}}{B^2 r_h^{6 z}+q_2^2 r_h^6+2 \alpha ^2 r_h^{z+7}},
\end{equation}
The expression for the Lorenz ratio is,
\begin{equation}
\bar{L}=\frac{ \bar{\kappa}_{xx}}{T \sigma_{xx}}=\frac{16 \pi ^2 r_h^{2 z+1} \left(B^2 r_h^{5 z}+\alpha ^2 r_h^7\right)}{\alpha ^2 \left(B^2 r_h^{6 z}+q_2^2 r_h^6+\alpha ^2 r_h^{z+7}\right)}.
\end{equation}

The explicit dependence of Lorenz ratio on the dynamical scaling and momentum relaxation strength is shown in Fig.17. 
\begin{figure}[h!]
  \includegraphics[width=.45\textwidth]{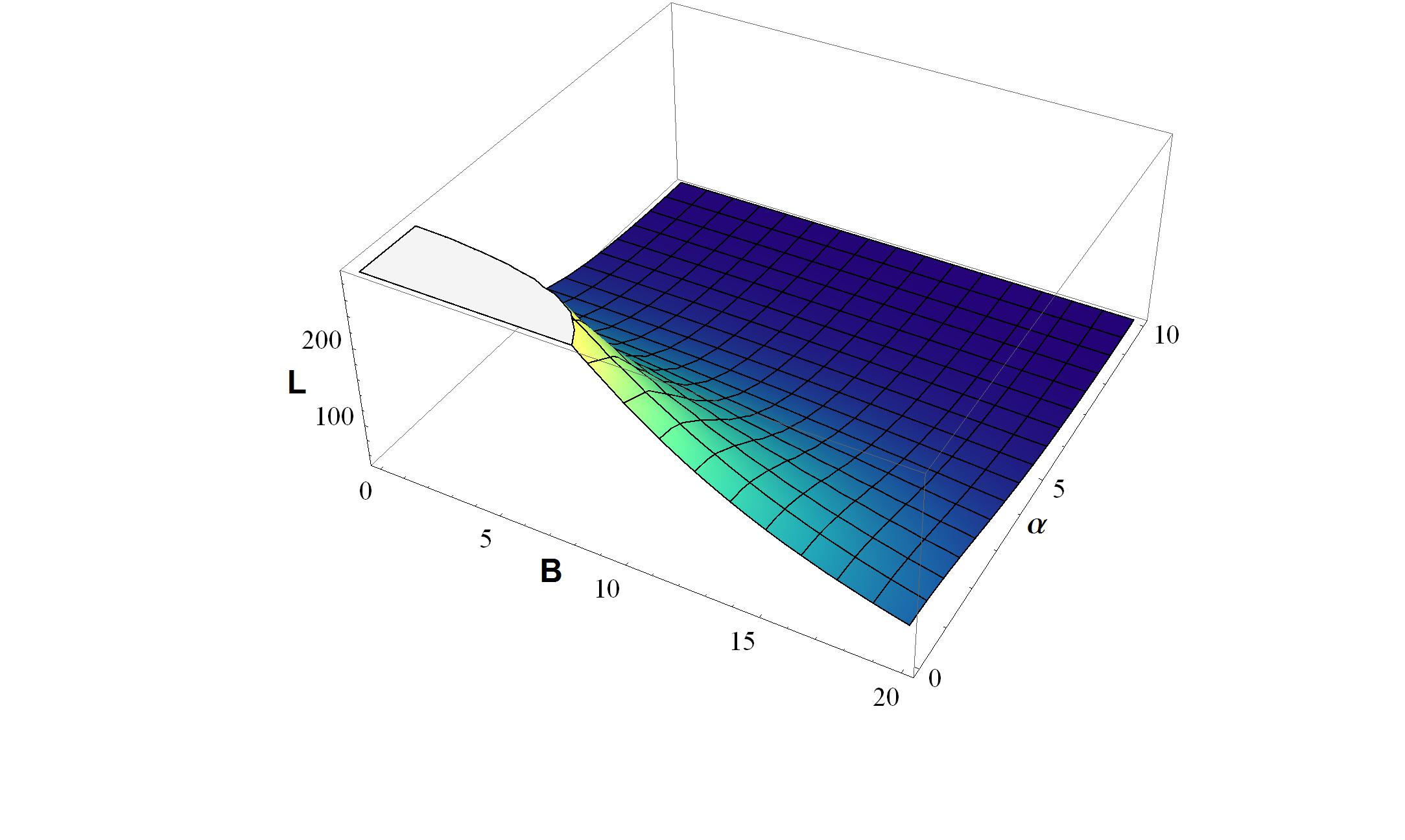}
  \includegraphics[width=.45\textwidth]{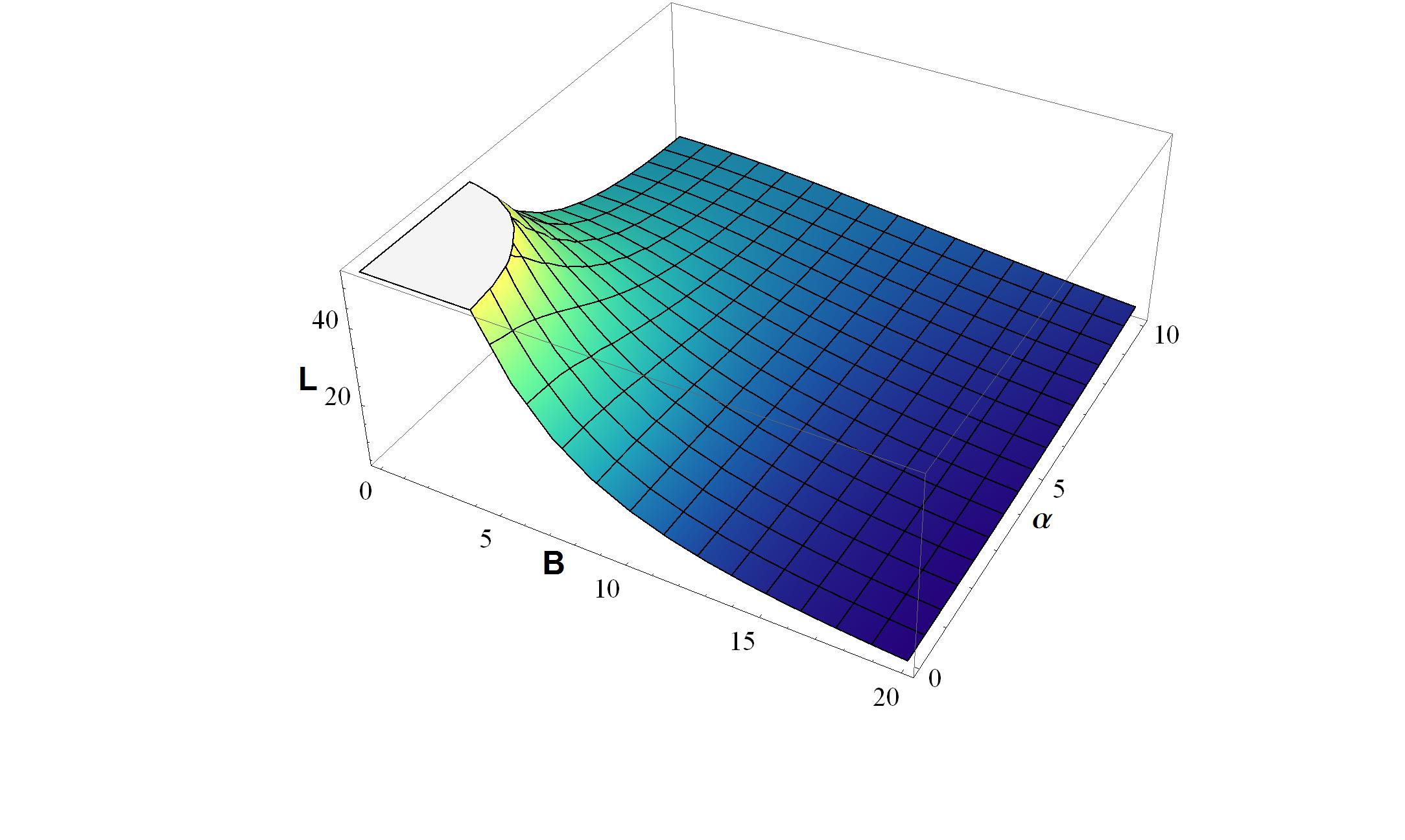}
\caption{Variation of  L with B and $\alpha$ for $z=1$(left) and $z=4/3$(right) }
\end{figure}

For $B \rightarrow 0$,
\begin{equation}
\bar{L}=\frac{16 \pi ^2 r_h^{2 z+2}}{q_2^2+\alpha ^2 r_h^{z+1}}.
\end{equation}
We obtain the Lorenz ratio ratio at zero temperature for vanishing magnetic field keeping $z=1$ as,
\begin{equation}\label{eq:L0}
\bar{L}=\frac{\bar{\kappa}_{xx}}{T \sigma_{xx}}|_{T, B \rightarrow 0}= \frac{4}{3} \pi ^2 \left(1+\frac{\alpha ^2}{\sqrt{\alpha ^4+12 q_2^2}}\right)
\end{equation}
The given expression indicates, at $B=0$ the WF law is valid and we obtain Fermi-liquid type ground state for $z=1$.

The temperature dependence of  Lorenz ratio is shown in Fig.18 for different momentum relaxation strength.  According to Wiedemann-Franz (WF) law the Lorenz ratio is constant for normal metals. Since our results show explicit dependence on temperature and also on Lifshitz scaling, the WF Law is violated in this model. The temperature dependence can not be extracted in a simple manner because of the interplay of different parameters in the system. We also plot the variation of the Lorenz ratio with the magnetic field in Fig.19. 

\begin{figure}[h!]
\centering
  \includegraphics[width=.38\textwidth]{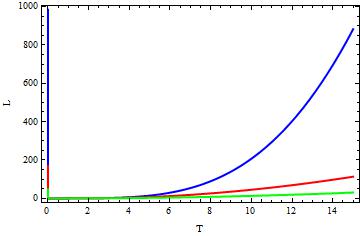}
\hspace{1cm}
  \includegraphics[width=.38\textwidth]{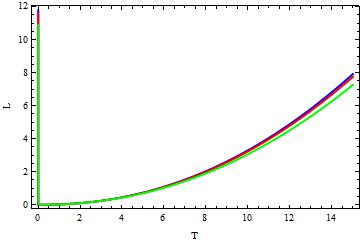}
\caption{L vs T at $\alpha$=0.1(blue),0.5(red),1(green) for $z=1$(left) and $z=4/3$(right) }
\end{figure}

\begin{figure}[h!]
\centering
  \includegraphics[width=.38\textwidth]{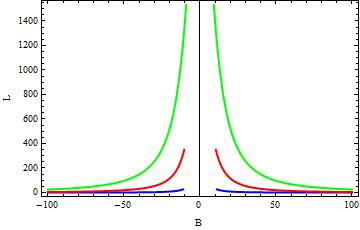}
\hspace{1cm}
  \includegraphics[width=.38\textwidth]{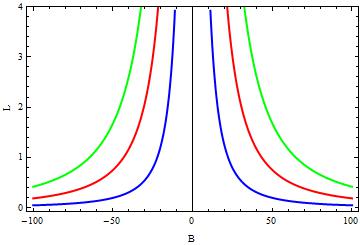}
\caption{L vs B at $T=$0.5(blue),1(red),1.5(green) for $z=1$(left) and $z=4/3$(right) }
\end{figure}

\section{Conclusions and Summary}
In this paper, we have investigated the DC transport of holographic systems with Lifshitz-like geometry and hyperscaling violation. The geometry is dual to non-relativistic ($z \neq 1$) condensed matter systems under the applied external magnetic field. We considered the near horizon limit of linearized equations of motion and calculated DC conductivity, thermoelectric and thermal conductivity analytically.  

We introduced the perturbations in both the gauge fields and obtained the expressions for the transport coefficients. The behavior of transport coefficients is depicted by numerical plots for different values, $z=1$ and $z=4/3$, of the dynamical exponent. While we have non-trivial scaling for $z=4/3$, the geometry reduces to RN-AdS black hole for $z=1$.  Dependence of DC transport on magnetic field and momentum relaxation strength is also studied while considering different Lifshitz scaling. The pattern of different type of conductivities ( $\sigma^{ij}_{xx} , \alpha^{ij}_{xx}$ and $\bar{\kappa}^{ij}_{xx}$) are quite similar showing monotonic dependence on the magnetic field but this behavior is absent in $\sigma^{ij}_{xy}$ etc . 

We also obtained the Hall angle, Seebeck coefficient and Lorenz ratio for the system and plotted them as a function of temperature and magnetic field. Hall angle shows $1/T^2$  in temperature behavior as given in equation (\ref{eq:ha1}) for $z=1$ but  it changes to  $1/T$ for $z=4/3$. The Wiedemann-Franz law is violated for our holographic model depicting unconventional metallic behavior. However, at zero temperature and $B \rightarrow 0$ the Fermi-liquid behavior is obtained for $z=1$.  Seebeck coefficient and Lorenz ratio also showed non-trivial dependence on the hyperscaling parameter. This unconventional dependence of transport coefficient on temperature can be useful to study the strange metal phenomenon \cite{Blake:2014yla,Amoretti:2016cad}.


\begin{thebibliography}{00}
\bibitem{Maldacena:1997re}
  J.~M.~Maldacena,
``The Large N limit of superconformal field theories and
supergravity,''
  Ad
v.\ Theor.\ Math.\ Phys.\  {\bf 2}, 231 (1998)
 [hep-th/9711200], 

\bibitem{Gubser:1998bc} 
  S.~S.~Gubser, I.~R.~Klebanov and A.~M.~Polyakov,
  ``Gauge theory correlators from noncritical string theory,''
  Phys.\ Lett.\ B {\bf 428}, 105 (1998)
  [hep-th/9802109].

\bibitem{Witten:1998qj} 
  E.~Witten,
  ``Anti-de Sitter space and holography,''
  Adv.\ Theor.\ Math.\ Phys.\  {\bf 2}, 253 (1998)
  [hep-th/9802150].

\bibitem{Son:2002sd}
 D.~T.~Son and A.~O.~Starinets,
``Minkowski space correlators in AdS / CFT correspondence: Recipe and applications,''  JHEP {\bf 0209}, 042 (2002)  
[hep-th/0205051].


\bibitem{Erlich:2005qh}
 J.~Erlich, E.~Katz, D.~T.~Son and M.~A.~Stephanov,
``QCD and a holographic model of hadrons,''
 Phys.\ Rev.\ Lett.\  {\bf 95}, 261602 (2005)
 [hep-ph/0501128],

\bibitem{DaRold:2005zs} 
  L.~Da Rold and A.~Pomarol,
 ``Chiral symmetry breaking from five dimensional spaces,''
 Nucl.\ Phys.\ B {\bf 721}, 79 (2005)
 [arXiv:0501218[hep-th]].
  
\bibitem{Karch:2006pv}
 A.~Karch, E.~Katz, D.~T.~Son and M.~A.~Stephanov,
 ``Linear confinement and AdS/QCD,''
  Phys.\ Rev.\ D {\bf 74}, 015005 (2006)
  [arXiv:0602229[hep-th]].


\bibitem{Hartnoll:2008kx} 
  S.~A.~Hartnoll, C.~P.~Herzog and G.~T.~Horowitz,
  JHEP {\bf 0812}, 015 (2008)
  [arXiv:0810.1563 [hep-th]].
    
\bibitem{Jain:2009bi}
 S.~Jain,
``Universal properties of thermal and electrical conductivity of gauge theory plasmas from holography,''
 JHEP {\bf 1006}, 023 (2010)
 [arXiv:0912.2719 [hep-th]].
    


\bibitem{Hartnoll:2008hs} 
  S.~A.~Hartnoll and C.~P.~Herzog,
  ``Impure AdS/CFT correspondence,''
  Phys.\ Rev.\ D {\bf 77}, 106009 (2008)
  [arXiv:0801.1693 [hep-th]].


\bibitem{Andrade:2013gsa} 
  T.~Andrade and B.~Withers,
  ``A simple holographic model of momentum relaxation,''
  JHEP {\bf 1405}, 101 (2014)
  [arXiv:1311.5157 [hep-th]].


\bibitem{Donos:2014yya} 
  A.~Donos and J.~P.~Gauntlett,
  ``The thermoelectric properties of inhomogeneous holographic lattices,''
  JHEP {\bf 1501}, 035 (2015)
  [arXiv:1409.6875 [hep-th]].


\bibitem{Amoretti:2014ola} 
  A.~Amoretti, A.~Braggio, N.~Magnoli and D.~Musso,
  ``Bounds on charge and heat diffusivities in momentum dissipating holography,''
  JHEP {\bf 1507}, 102 (2015)
  [arXiv:1411.6631 [hep-th]].

\bibitem{Davison:2013jba} 
  R.~A.~Davison,
  ``Momentum relaxation in holographic massive gravity,''
  Phys.\ Rev.\ D {\bf 88}, 086003 (2013)
  [arXiv:1306.5792 [hep-th]].


\bibitem{Blake:2013bqa} 
  M.~Blake and D.~Tong,
  Phys.\ Rev.\ D {\bf 88}, no. 10, 106004 (2013)
  [arXiv:1308.4970 [hep-th]].

\bibitem{Vegh:2013sk} 
  D.~Vegh,
  ``Holography without translational symmetry,''
 [ arXiv:1301.0537 [hep-th]].

\bibitem{Amoretti:2014mma} 
  A.~Amoretti, A.~Braggio, N.~Maggiore, N.~Magnoli and D.~Musso,
  ``Analytic dc thermoelectric conductivities in holography with massive gravitons,''
  Phys.\ Rev.\ D {\bf 91}, no. 2, 025002 (2015)
  [arXiv:1407.0306 [hep-th]].

\bibitem{Baggioli:2014roa} 
  M.~Baggioli and O.~Pujolas,
 ``Electron-Phonon Interactions, Metal-Insulator Transitions, and Holographic Massive Gravity,''
  Phys.\ Rev.\ Lett.\  {\bf 114}, 25 (2015)
  [arXiv:1411.1003 [hep-th]].


\bibitem{Kachru:2008yh} 
  S.~Kachru, X.~Liu and M.~Mulligan,
 ``Gravity duals of Lifshitz-like fixed points,''
  Phys.\ Rev.\ D {\bf 78}, 106005 (2008)
  [arXiv:0808.1725 [hep-th]].

\bibitem{Liu:2009dm} 
  H.~Liu, J.~McGreevy and D.~Vegh,
  ``Non-Fermi liquids from holography,''
  Phys.\ Rev.\ D {\bf 83}, 065029 (2011)
  [arXiv:0903.2477 [hep-th]].

\bibitem{Bertoldi:2010ca} 
  G.~Bertoldi, B.~A.~Burrington and A.~W.~Peet,
  ``Thermal behavior of charged dilatonic black branes in AdS and UV completions of Lifshitz-like geometries,''
  Phys.\ Rev.\ D {\bf 82}, 106013 (2010)
  [arXiv:1007.1464 [hep-th]].

\bibitem{Berglund:2011cp} 
  P.~Berglund, J.~Bhattacharyya and D.~Mattingly,
  ``Charged Dilatonic AdS Black Branes in Arbitrary Dimensions,''
  JHEP {\bf 1208}, 042 (2012)
  [arXiv:1107.3096 [hep-th]].


\bibitem{Iizuka:2011hg} 
  N.~Iizuka, N.~Kundu, P.~Narayan and S.~P.~Trivedi,
  ``Holographic Fermi and Non-Fermi Liquids with Transitions in Dilaton Gravity,''
  JHEP {\bf 1201}, 094 (2012)
  [arXiv:1105.1162 [hep-th]].

\bibitem{Gouteraux:2011ce} 
  B.~Gouteraux and E.~Kiritsis,
  ``Generalized Holographic Quantum Criticality at Finite Density,''
  JHEP {\bf 1112}, 036 (2011)
  [arXiv:1107.2116 [hep-th]].



\bibitem{Alishahiha:2012cm} 
  M.~Alishahiha and H.~Yavartanoo,
  ``On Holography with Hyperscaling Violation,''
  JHEP {\bf 1211}, 034 (2012)
  [arXiv:1208.6197 [hep-th]].

\bibitem{Dey:2012tg} 
  P.~Dey and S.~Roy,
  ``Lifshitz-like space-time from intersecting branes in string/M theory,''
  JHEP {\bf 1206}, 129 (2012)
  [arXiv:1203.5381 [hep-th]].

\bibitem{Alishahiha:2012qu} 
  M.~Alishahiha, E.~O Colgain and H.~Yavartanoo,
  ``Charged Black Branes with Hyperscaling Violating Factor,''
  JHEP {\bf 1211}, 137 (2012)
  [arXiv:1209.3946 [hep-th]].



\bibitem{Gouteraux:2014hca} 
  B.~Goutéraux,
  ``Charge transport in holography with momentum dissipation,''
  JHEP {\bf 1404}, 181 (2014)
  [arXiv:1401.5436 [hep-th]].

\bibitem{Cremonini:2016avj} 
  S.~Cremonini, H.~S.~Liu, H.~Lu and C.~N.~Pope,
  ``DC Conductivities from Non-Relativistic Scaling Geometries with Momentum Dissipation,''
  JHEP {\bf 1704}, 009 (2017)
  [arXiv:1608.04394 [hep-th]].

\bibitem{Karch:2014mba} 
  A.~Karch,
  ``Conductivities for Hyperscaling Violating Geometries,''
  JHEP {\bf 1406}, 140 (2014)
  [arXiv:1405.2926 [hep-th]].

\bibitem{Charmousis:2010zz} 
  C.~Charmousis, et.al. 
  ``Effective Holographic Theories for low-temperature condensed matter systems,''
  JHEP {\bf 1011}, 151 (2010)
  [arXiv:1005.4690 [hep-th]].

\bibitem{Ammon:2012je} 
  M.~Ammon, M.~Kaminski and A.~Karch,
 ``Hyperscaling-Violation on Probe D-Branes,''
  JHEP {\bf 1211}, 028 (2012)
  [arXiv:1207.1726 [hep-th]].


\bibitem{Kuang:2014pna} 
  X.~M.~Kuang, E.~Papantonopoulos, B.~Wang and J.~P.~Wu,
  ``Formation of Fermi surfaces and the appearance of liquid phases in holographic theories with hyperscaling violation,''
  JHEP {\bf 1411}, 086 (2014)
  [arXiv:1409.2945 [hep-th]],


\bibitem{Kuang:2015mlf} 
  X.~M.~Kuang and J.~P.~Wu,
 ``Transport coefficients from hyperscaling violating black brane: shear viscosity and conductivity,''
  arXiv:1511.03008 [hep-th].


\bibitem{Wu:2015ajt} 
  J.~P.~Wu and X.~M.~Kuang,
 ``Scalar Boundary Conditions in Hyperscaling Violating Geometry,''
  Phys.\ Lett.\ B {\bf 753}, 34 (2016)
  [arXiv:1512.03499 [hep-th]].


\bibitem{Li:2016rcv} 
  L.~Li,
  ``Hyperscaling Violating Solutions in Generalised EMD Theory,''
  Phys.\ Lett.\ B {\bf 767}, 278 (2017)
  [arXiv:1608.03247 [hep-th]].


\bibitem{Amoretti:2015gna} 
  A.~Amoretti and D.~Musso,
  ``Magneto-transport from momentum dissipating holography,''
  JHEP {\bf 1509}, 094 (2015)
  [arXiv:1502.02631 [hep-th]].


\bibitem{Donos:2015bxe} 
  A.~Donos, J.~P.~Gauntlett, T.~Griffin and L.~Melgar,
  ``DC Conductivity of Magnetised Holographic Matter,''
  JHEP {\bf 1601}, 113 (2016)
  [arXiv:1511.00713 [hep-th]].


\bibitem{Kim:2015wba} 
  K.~Y.~Kim, K.~K.~Kim, Y.~Seo and S.~J.~Sin,
  ``Thermoelectric Conductivities at Finite Magnetic Field and the Nernst Effect,''
  JHEP {\bf 1507}, 027 (2015)
  [arXiv:1502.05386 [hep-th]].

\bibitem{Blake:2015ina} 
  M.~Blake, A.~Donos and N.~Lohitsiri,
  ``Magnetothermoelectric Response from Holography,''
  JHEP {\bf 1508}, 124 (2015)
  [arXiv:1502.03789 [hep-th]].

\bibitem{Papadimitriou:2004rz} 
  I.~Papadimitriou and K.~Skenderis,
  ``Correlation functions in holographic RG flows,''
  JHEP {\bf 0410}, 075 (2004)
  [hep-th/0407071].

\bibitem{Iqbal:2008by} 
  N.~Iqbal and H.~Liu,
  ``Universality of the hydrodynamic limit in AdS/CFT and the membrane paradigm,''
  Phys.\ Rev.\ D {\bf 79}, 025023 (2009)
  [arXiv:0809.3808 [hep-th]]

\bibitem{Kuperstein:2011fn} 
  S.~Kuperstein and A.~Mukhopadhyay,
  ``The unconditional RG flow of the relativistic holographic fluid,''
  JHEP {\bf 1111}, 130 (2011)
  [arXiv:1105.4530 [hep-th]].
  
\bibitem{Heemskerk:2010hk}
        I.~Heemskerk and J.~Polchinski,
        ``Holographic and Wilsonian Renormalization Groups,''
        JHEP {\bf 1106}, 031 (2011)
        [arXiv:1010.1264 [hep-th]].

\bibitem{Faulkner:2010jy} 
  T.~Faulkner, H.~Liu and M.~Rangamani,
  ``Integrating out geometry: Holographic Wilsonian RG and the membrane paradigm,''
  JHEP {\bf 1108}, 051 (2011)
  [arXiv:1010.4036 [hep-th]].

\bibitem{Kuperstein:2013hqa} 
  S.~Kuperstein and A.~Mukhopadhyay,
  ``Spacetime emergence via holographic RG flow from incompressible Navier-Stokes at the horizon,''
  JHEP {\bf 1311}, 086 (2013)
  [arXiv:1307.1367 [hep-th]].

\bibitem{Amoretti:2014zha} 
  A.~Amoretti, A.~Braggio, N.~Maggiore, N.~Magnoli and D.~Musso,
  ``Thermo-electric transport in gauge/gravity models with momentum dissipation,''
  JHEP {\bf 1409}, 160 (2014)
  [arXiv:1406.4134 [hep-th]].

\bibitem{Ge:2016lyn} 
  X.~H.~Ge et.al.
  ``Linear and quadratic in temperature resistivity from holography,''
  JHEP {\bf 1611}, 128 (2016)
  [arXiv:1606.07905 [hep-th]].

\bibitem{Tian:2017vfk} 
  Y.~Tian, X.~H.~Ge and S.~F.~Wu,
  ``Wilsonian RG-flow approach to holographic transport with momentum dissipation,''
  Phys.\ Rev.\ D {\bf 96}, no. 4, 046011 (2017)
  [arXiv:1702.05470 [hep-th]].

\bibitem{Hartnoll:2007ai} 
  S.~A.~Hartnoll and P.~Kovtun,
  ``Hall conductivity from dyonic black holes,''
  Phys.\ Rev.\ D {\bf 76}, 066001 (2007)
  [arXiv:0704.1160 [hep-th]].

\bibitem{Hartnoll:2007ih} 
  S.~A.~Hartnoll et.al. 
  ``Theory of the Nernst effect near quantum phase transitions in condensed matter, and in dyonic black holes,''
  Phys.\ Rev.\ B {\bf 76}, 144502 (2007)
  [arXiv:0706.3215 [cond-mat.str-el]], Zhang, Y. et. al.,``Determining the Wiedemann-Franz Ratio from the Thermal Hall Conductivity: Application to Cu and $ YBa_{2}Cu_{3}{O}_{6.95}$", Phys. Rev. Lett.,{\bf 84},10 (2000),[arXiv:cond-mat/0001037].


\bibitem{Blake:2014yla} 
  M.~Blake and A.~Donos,
 ``Quantum Critical Transport and the Hall Angle,''
  Phys.\ Rev.\ Lett.\  {\bf 114}, no. 2, 021601 (2015)


\bibitem{Ge:2016sel} 
  X.~H.~Ge, Y.~Tian, S.~Y.~Wu and S.~F.~Wu,
  ``Anomalous transport of the cuprate strange metal from holography,''
  [arXiv:1606.05959 [hep-th]].

\bibitem{Amoretti:2017xto} 
  A.~Amoretti, A.~Braggio, N.~Maggiore and N.~Magnoli,
  ``Thermo-electric transport in gauge/gravity models,''
  Adv.\ Phys.\ X {\bf 2}, no. 2, 409 (2017).


\bibitem{Cremonini:2017qwq} 
  S.~Cremonini, A.~Hoover and L.~Li,
  ``Backreacted DBI Magnetotransport with Momentum Dissipation,''
[arXiv:1707.01505 [hep-th]].

\bibitem{Amoretti:2016cad} 
  A.~Amoretti et.al.
  ``Chasing the cuprates with dilatonic dyons,''
  JHEP {\bf 1606}, (2016).
  [arXiv:1603.03029 [hep-th]].


\bibitem{Matsuo:2011fk} 
  Y.~Matsuo, S.~J.~Sin and Y.~Zhou,
  ``Mixed RG Flows and Hydrodynamics at Finite Holographic Screen,''
  JHEP {\bf 1201}, 130 (2012)
  [arXiv:1109.2698 [hep-th]].

\bibitem{Andri:1997}
P.~W.~Anderson,
``The theory of superconductivity in the high $T_c$ cuprates",
Princeton University Press, Princeton (1997).

\bibitem{Hussey:2008}
N.E. Hussey,
`` Phenomenology of the normal state in-plane transport properties of high-Tc cuprates,"
Journal of Physics: Condensed Matter, (2008).

\bibitem{Kim:2004}
``Thermoelectric power of $La_{2-x}Sr_xCuO_4$ at high temperatures'',
Ann. Phys. (13) 43  (2004).

\end{thebibliography}
\end{document}